# Transcriptome Complexities Across Eukaryotes

JAMES E. TITUS-MCQUILLAN[1*], ADALENA V. NANNI[2,3], LAUREN M. MCINTYRE[2,3], & REBEKAH L. ROGERS[1]

[1]Department of Bioinformatics and Genomics, University of North Carolina at Charlotte, Charlotte, NC 28223
[2]Department of Molecular Genetics and Microbiology, University of Florida, Gainesville, FL 32611
[3]University of Florida Genetics Institute, University of Florida, Gainesville Florida 32611
[*]Corresponding Author: JAMES E. TITUS-MCQUILLAN

## Abstract:

Genomic complexity is a growing field of evolution, with case studies for comparative evolutionary analyses in model and emerging non-model systems. Understanding complexity and the functional components of the genome is an untapped wealth of knowledge ripe for exploration. With the "remarkable lack of correspondence" between genome size and complexity, there needs to be a way to quantify complexity across organisms. In this study we use a set of complexity metrics that allow for evaluation of changes in complexity using TranD. We ascertain if complexity is increasing or decreasing across transcriptomes and at what structural level, as complexity is varied. We define three metrics – TpG, EpT, and EpG in this study to quantify the complexity of the transcriptome that encapsulate the dynamics of alternative splicing. Here we compare complexity metrics across 1) whole genome annotations, 2) a filtered subset of orthologs, and 3) novel genes to elucidate the impacts of ortholog and novel genes in transcriptome analysis. We also derive a metric from Hong et al., 2006, Effective Exon Number (EEN), to compare the distribution of exon sizes within transcripts against random expectations of uniform exon placement. EEN accounts for differences in exon size, which is important because novel genes differences in complexity for orthologs and whole transcriptome analyses are biased towards low complexity genes with few exons and few alternative transcripts. With our metric analyses, we are able to implement changes in complexity across diverse lineages with greater precision and accuracy than previous cross-species comparisons under ortholog conditioning. These analyses represent a step forward toward whole transcriptome analysis in the emerging field of non-model evolutionary genomics, with key insights for evolutionary inference of complexity changes on deep timescales across the tree of life. We suggest a means to quantify biases generated in ortholog calling and correct complexity analysis for lineage-specific effects. With these metrics, we directly assay the quantitative properties of newly formed lineage-specific genes as they lower complexity in transcriptomes.

# Introduction:

Transcriptome complexity is the product of evolutionary, biophysical, and molecular constraints that depend on the environmental context of an organism (Lynch & Conery, 2003). For each gene, alternative splicing allows more diverse variations of proteins from a single DNA locus, increasing the number of structures that can arise from a given genomic region (Gilbert, 1978; C. elegans Sequencing Consortium*, 1998; International Human Genome Sequencing Consortium, 2004). Moreover, the combination of different exons and splicing action may change regulatory profiles for transcript/protein produced (Black, 2000). This isoform complexity increases information content stored in a single genomic region (Kwan et al., 2008) often with functional consequences (Kalsotra & Cooper, 2011; Kragh-Hansen et al., 2013). The ways that global genome-wide transcript structures change across species influence the scope and consequences of genetic diversity and genetic architecture (Nilsen & Graveley, 2010).

Alternative splicing is a distinct but pervasive product of Eukaryotic evolution. Among different organisms, patterns of alternative splicing and transcriptome complexity may shift, producing differences in exon/intron boundaries, length, and number (Furlanis & Scheiffele, 2018). Some of these changes are linked directly to genetic and biochemical changes in spliceosome machinery including U11 and U12 spliceosomes (Moyer et al. 2020), long intron splicing dynamics (Shepard et al., 2009), and specific tissue regulation (Sánchez, 2004; Katayama et al., 2005). Others may be the product of selection for specific genetic features and functions within the cell (Hodges & Bernstein, 1994; Matlin et al., 2005; Schad et al., 2011). Complexity is driven by many factors such as tissue type (Sealfon et al., 2021), sex-specificity (Maine et al., 1985; Schutt & Nothiger, 2000), development (Sánchez, 2004; Nilsen & Graveley, 2010), phenotype (Bonner, 1988), and biochemical constraints (Qian, 2009). These factors are associated with multiple modes of genetic evolution across the tree of life including sex with reproduction, body formation and tissue complexity, and pathogens/symbioses (Holland, 1999; Adami, 2002). Intron density and distribution can vary by orders of magnitude across unicellular and multicellular taxa (Roy & Irimia, 2009; Curtis et al., 2012), sometimes including complete spliceosome loss or reemergence of splicing function (Irimia & Roy, 2014). Modes of exon use, levels of alternative splicing, and isoforms generated may change as spliceosome structures evolve. Given the pervasiveness and consequences of alternative splicing, analysis of genome-wide patterns for alternative splicing may reveal biological variation that shapes molecular and evolutionary biology.

One of the challenges of comparative transcriptome and splicing analysis across the tree of life has been the (seemingly) "necessary evil" of assigning orthology across distantly related taxa (Holland, 1999; Shimeld & Holland, 2000; Rogozin et al., 2003; Kumar, 2009; Irimia & Roy, 2014). Over time, the evolution of novel genes results in lower concordance in gene content across organisms. Heterogeneity in gene content has been highlighted in phylogenetics as an analytical complication, but it is also a biological contributor to changes in species complexity (Bravo et al. 2019; Smith & Hahn, 2021). Novel genes are associated with fewer numbers of exons, shorter exons, and lower biochemical complexity, along with increasing sequence ambiguity (Long, 2001; Kaessmann, 2010; Willis & Masel, 2018). Failure to include novel genetic elements may therefore alter estimates of complexity as a genomic trait. Moreover, as genetic distance increases, sequence alignment and ortholog calling becomes more challenging when generating additional sources of uncertainty and potential bias (Emms & Kelly, 2015). As a result of these many analytical complications, studies of intron-exon placement, complexity, and splicing patterns with ortholog conditioning may be more affected across vast evolutionary distance. Using subsets of constrained genome sequences may portray evolutionary processes for single genes that disagree with the evolutionary history of the species due to incomplete lineage sorting (ILS), introgression, and gene duplication and loss (GDL) (Maddison, 1997). Sets of proteins that are highly conserved across distantly related taxa may make inference simpler (Fitch, 1970; Day, 1983; De Queiroz & Donoghue, 1988), but selections of specific proteins may introduce biases against rapidly evolving genes (Dover, 1987; Koonin, 2005; Ellegren & Parsch, 2007; O'Toole et al., 2018; Begum & Robinson-Rechavi, 2021).

In the face of these challenges, clear solutions for complexity analysis are needed that are robust and free from biases of ortholog conditioning. While single gene analysis on close evolutionary distance may be best served by the precision of cross-species alignment of introns and exons, whole genome transcriptome splicing may be assayed as a global phenotype that can shift across species. Comparing complexity within a single species' transcriptome, we can estimate the properties of all genes at once in an ortholog-free framework. These global, whole-genome properties may provide more complete information about transcriptome architecture than previous approaches comparing gene-by-gene. A new program for transcriptome analysis, TranD (Nanni, et al. doi: https://doi.org/10.1101/2021.09.28.462251; https://github.com/McIntyre-Lab/TranD), enables comparisons of complexity across a wide array of genomes across the eukaryotic tree of life. We define complexity with these principles in mind and quantify complexity metrics from TranD here as variation among genetic regions with respect to unique exons per gene (EpG), exons per transcript (EpT), and transcripts per gene (TpG). Additionally, we add one previously developed measure of transcriptome complexity, the effective exon number (EEN). In alternative splicing, EpG defines the total number of genetic elements available to generate unique combinations, while TpG is the product of the total of these combinations. EpT governs the number of introns that must be spliced, influencing transcript regulation. Finally, EEN is the product of splice junction spacing reflecting biochemical constraints of splicing factors. Using these metrics, we quantify complexity across phylogenies for major metazoan taxa: Deuterostomes, *Drosophila*, Plants, and Fungi. We focus on these well-annotated taxa as examples for how evolutionary analysis of transcriptome complexity may guide inference of evolutionary processes.

The analyses presented here offer an important case study for comparative evolutionary analyses in model and emerging non-model systems. Quantifying complexity as a set of metrics allows for evaluation of changes in complexity, ascertain if complexity is increasing or decreasing across transcriptomes and at what structural level. Give that there are a variety of levels to complexity, one metric does not fit all and present three distinct complexity metrics defined above. Here we compare results from 1) whole genome annotations, 2) a filtered subset of orthologs, and 3) novel genes to elucidate the impacts of orthology and novel genes in transcriptome analysis. We suggest a means to quantify biases generated in ortholog calling and correct complexity analysis for lineage-specific effects. With these metrics, we directly assay the quantitative properties of newly formed lineage-specific genes as they lower complexity in transcriptomes. We implement evolutionary rate analyses from complexity changes across diverse lineages with greater precision and accuracy than previous cross-species comparisons under ortholog conditioning. These analyses represent a step forward toward whole transcriptome analysis in the emerging field of non-model evolutionary genomics, with key insights for evolutionary inference of complexity changes on deep timescales across the tree of life.

## Methods:
*TranD complexity calculations*
We used the new transcriptome analysis software TranD to calculate three metrics of transcriptome complexity within species: TpG, EpT, and EpG (Nanni et al., https://doi.org/10.1101/2021.09.28.462251; https://github.com/McIntyre-Lab/TranD/wiki). The exons per gene (EpG) metric was initially derived from Spieth & Lawson, (2005). These three generalized metrics describe the global phenotypic structure of transcriptomes and are used to quantify transcriptome complexity, from annotations, of lineages across the tree of life. TranD derives these metrics by consolidating exons and transcripts into their parent genes (for both exons and transcripts) and into specific transcripts (for exons only) to illustrate the dynamics of between various coordinate systems, transcriptome structure, and alternative splicing (AS) among and between lineages. We used describe_transcriptome_complexity_GTF.py script to quantify transcriptome complexity for each organism using GTF files obtained from NCBI RefSeq and GeneBank. Each organism was then consolidated to one file, per each group of interest [Deuterostome, *Drosophila*, Plantae, and Fungi], using merge_species_transcriptome_info_counts.py script. Complexities were described for partitions for orthologs filters and novel gene filters (see ortholog identification).

First we tested for normality using the Shapiro-Wilk Normality Test using the shapiro.test() function in R (Royston, 1982). To evaluate the interdependence between our metrics in the data we present, we used the Pearson correlation statistic using the lm() function in R (Wilkinson & Rogers, 1973; Chambers, 1992). If the data set was normally distributed, we used the Pearson correlation, if it was not the Spearman's rank correlation coefficient was used for nonparametric data (Supplementary Information).

*Effective Exon Number*
We estimated the "Effective Exon Number" for each transcript according to models previously developed by Hong et al. (2006). For each transcript, the Effective Exon Number (EEN, formerly reported as $N_e$) is given by $EEN = 1/(\sum_{i=1}^{EpT} (1/(L_e^2)))$ where *i* goes from 1 to EpT, the total number of Exon per Transcript, and $L_e$ is the exon length scaled to a proportion of total transcript length. EEN is naturally bounded by [0, EpT].

EEN depends directly on the distribution of exon lengths and the dispersion or clustering of intron positions in cDNA transcripts. EEN is equal to the number of exons (EEN = EpT) if a transcript contains evenly spaced introns (overdispersion) with equal exon lengths throughout the transcript. Lower values (EEN << EpT) represent more clustered intron positions and under dispersed distribution of exon lengths (Hong et al. 2006). Theoretical predictions would suggest that exon fragment lengths should follow a Broken Stick model (Holst 1980). Where the model takes unit of length and randomly (and simultaneously) selecting break points from a uniform distribution breaking it into *N* pieces. Values above the Broken Stick model suggest biological or analytical factors that create more evenly spaced intron breaks than the null. Values below this null model suggest factors that create more tightly clustered distributions of intron breakpoints across the transcript.

*Ortholog identification*
We used TranD to analyze transcriptomes for test clades of Eukaryotes where reference genomes and whole transcriptome annotations were available in RefSeq and OrthoDB. Organism annotation data was selected by covering tractable data from well-studied phylogenetic lineages in the Eukarya tree of life as possible. Reference files were procured from the Refseq database in GTF format (http://www.ncbi.nlm.nih.gov/refseq/). Guidelines for proper annotation acquisition are as follows:

i) Annotations must be available on RefSeq and OrthoDB 10v1 (or v9.1 for *Drosophila* only). RefSeq is a standardized public database that is actively curated providing the most comprehensive and rich annotations of the tree life available. This allows for the highest quality annotation among lineages to pull from in our analyses.

ii) The reference organisms must be annotated and have an assembled and annotated reference genome. Again, to facilitate the best annotated sequence products available within lineages.

iii) The reference must have an NCBI release. All RefSeq genome annotations have an NCBI release, meaning they were processed by biological experts using the RefSeq processing pipeline (Pruitt *et al.*, 2009; O'Leary et al., 2016).

Complete references were gathered where available. Some key lineages are not available on RefSeq at high quality and so were omitted, e.g., Myxini, Tardigrada, Onychophora, *etc*. For some organisms the Fungi group only GenBank (GCA) reference (n = 16/77) were available. To have the adequate phylogenetic representation required for robust estimation, we choose to use GCA references for some constituents with n = 61/77 having RefSeq releases (GCF).

Orthologs were gathered from the OrthoDB v10.1 portal, where taxon organism IDs were collected. Orthologs were considered by selecting the most recent common ancestor (MRCA) for the groups (level on OrthoDB) and follow a 1:Multiple selection of orthologs. Input files needed to parse novel and orthologous genes, downloaded from OrthoDB's data section (https://www.orthodb.org/?page=filelist), include <odb10v1_OG2genes.tab.gz> and <odb10v1_gene_xrefs.tab.gz>. Our code translates ortholog group IDs from OrthoDB to xrefs and parses NCBI reference GTFs generating a total of three GTF files – [original] whole-transcriptome GTF, ortholog GTF, and novel gene GTF. Ortholog selection was conducted through a series of python scripts to filter OrthoDB orthologs from novel lineage specific genes NCBI GTF files. Full descriptions with examples for using code can be found in on GitHub: https://github.com/jemcquillan/OrthoDB_Parser.

To estimate whether there are significant differences between the whole-transcriptome, orthologous, and novel genetic data we ran a two-sample Wilcoxon Rank Sum (Mann-Whittney Test) using the wilcoxon.test() function in R (Bauer. 1972; Hollander & Wolfe, 1973). Estimated p-values were adjusted with Bonferroni using the p.adjust() function in R (Bland & Altman, 1995). We estimated if there are biases in complexity metric variances between orthologs and whole-transcriptomes by cross-validating with a Jackknife resampling with 10,000 replicates to ensure that differences ortholog subsets of the transcriptome are not biased by lack of independence. Given that the distributions of EpT, EpG, and TpG are independent per organism, a Fisher's combined probability test was conducted for each complexity metric using the R package poolr's fisher() function (Fisher, 1932; Brown, 1975; Higham, 2002; Cinar & Viechtbauer, 2022).

*Evolutionary rate analysis*
If complexity metrics differ for genes that have orthologs across phylogenies compared with complex datasets that include lineage specific genes, conditioning on orthology could introduce biases in evolutionary rate analysis of transcriptomes (Roy & Irimia, 2009; Irimia & Roy, 2014). To understand how ortholog complexities behave compared to whole transcriptome complexities in downstream analyses, we ran evolutionary rates on complexity traits estimated using whole transcriptome data and conditioning on genes with orthologs across the entire phylogeny. We compare evolutionary rates using PhyTools (Revell, 2012) and BAMM (Rabosky, 2014) to determine whether there are significant differences when conditioning on the presence of orthologs across taxa. We used PhyTools ratebytree() function (O'Meara et al., 2006; Adams, 2012; Revell, 2012; Revell et al. 2018) to ascertain if the evolutionary rate between orthologs and whole-transcriptome data has any bearing on which dataset to use for a specific group. BAMM was used to find mean phylorate shifts across both whole-transcriptome complexity metrics and orthologous gene complexity metrics to compare variation in complexity rate shifts across the phylogeny. This was conducted to understand the dynamics between complexity between whole-transcriptome annotations vs. annotations conditioned on orthologous genes.

Phylogenetic analysis was calibrated using divergence times from the Time Tree of Life (TToL) (Hedges et al., 2006; Kumar et al., 2017). The TToL constructs phylogenetic relationships through meta-analysis of currently published time-calibrated phylogenies. Each broad group (Deuterostomia, Drosophila, Plantae, and Fungi) had a phylogeny constructed from the TimeTree portal. If the TimeTree database did not have a specific individual that had criteria for annotation for selection, we picked closely related sister taxa present in the tree as comparable divergence times at the same relative tip placement of the phylogeny. For Fungi, phylogenies are still on going for proper placement of many taxa, so metrics were consolidated by class, the taxonomic grouping currently offered in the TimeTree database.

We used Phytools ratebytree() function for continuous traits, using the "OU" model of trait evolution to see if the rate is equal among all trees, or if the rates or regimes can differ between trees. The phylogenies were identical for each group. We used mean TpG, EpT, and EpG independently as traits. Then having the evolutionary rate estimated by complexity trait for ortholog metrics and whole transcriptome complexities. The posthoc() function (Revell, 2012) in PhyTools was then conducted to test if there was a significant difference between orthologous portioned complexity metrics compared to whole transcriptome complexity metrics.

BAMM was run across 10,000,000 generations with a sampling frequency every 1000 Markov Chains. Burnins, for all groups, were set to a 10% burnin (a total of 1000 Markov Chains).

Plotting of the BAMM and downstream analyses were conducting using BAMMTools R package (Rabosky et al., 2014). We checked effective sampling sizes (ESS) and convergence of MCMC runs. All our runs ran to convergence, with ESS > 200 in most cases. To assess if the ESS numbers could be higher, we ran multiple runs with longer generate time, and combined runs. Here the ESS numbers did not change, yet still ran to convergence. Given our analysis did converge and ESS numbers did not dramatically change, we proceeded given the data at hand. Mean phylorate shift was calculated from the getEventData() function in BAMMTools R package. Rate shift probabilities were gathered using BAMMTools built-in functions to extract posterior-probabilities of a species [getTipRates()], a monophyletic gorup [getCladeRates()], or a branch in the phylogeny [getMarginalBranchRateMatrix()].

## Results:
*Complexity metrics*
TranD empirically estimates complexity by calculating metrics that are immediately comparable across species with rigorous and repeatable analytical criteria (Nanni, et al. doi: https://doi.org/10.1101/2021.09.28.462251). Each species is estimated independently from all others, generating complexity metrics for each transcriptome. These metrics offer complexity measures as genetic traits that can be compared across a phylogeny. Because each complexity is derived from a single branch in the tree of life, there is breadth to compare deep-time divergence between and among distantly related species. We use four major clades as test cases to assay genome complexity across well annotated phylogenies: Deuterostomia, *Drosophila, Plantae*, and Fungi. Each of these clades spans different timescales (from ~50 mya in the *Drosophila* group to ~1 billion years ago in the fungi group) and unique biological features. These examples show how analysis of transcriptome complexity can clarify unique genetic properties of species within clades. We show how such inference is affected by orthology (and genetic novelty) across highly divergent phylogenetic groups.

Transcriptome complexities vary across organisms from distantly related lineages (Fig. 1). Among the deuterostome clade mean TpG is between 1.00-3-97. The means between EpT and EpG are more concordant with mean EpT 7.10-13.31 and mean EpG is 7.10-11.09. Another well-studied metazoan group is the genus *Drosophila*. *Drosophila* have smaller transcript complexity metrics than those of deuterostome lineages. This is concordant with genome size disparities between the two groups. The range of EpT across species are 3.38-6.08 and the means of EpG are 3.38-5.01, which do not overlap deuterostome EpT and EpG mean metrics. TpG means are lower too at 1.00-1.97 TpG. Among taxa in the Plantae division Viridiplantae the mean TpG range is 1.19-2.39, mean EpT range is 4.76-8.14, and mean EpG range is 4.32-8.25. Fungi lineages have a much lower distribution in complexity compared to the other higher-level taxonomic lineages. Fungi mean TpG range is 1.00-1.29. Fungi EpT mean ranges from 1.03-6.62. Finally, the Fungi mean EpG is of 1.03-6.63. Across all higher lineages: Deuterostomia, Drosophila, Plantae, and Fungi, mean TpG is significantly different across groups, KW *Chi*$^2$ = 148.4, P = 5.777e-32. Both EpT and EpG too are significantly different across higher lineages, for EpT KW *Chi*$^2$ = 160.9, P = 1.19e-34 and EpG KW *Chi*$^2$ = 155.1, P = 2.08e-33 (Supp. Mat. Mean Complexity Metrics.pdf). Hence, each clade portrays a characteristic complexity that is the unique product of biochemical, molecular, and evolutionary properties of the organisms. See Supp. Fig. 1 for a complete list of complexity metrics across each organism.

*Complexity Whole-transcriptome and Orthologs*
Orthologous sequences are shared genetic code between species that share an ancestor, separated by a speciation event (Finch, 1970). Orthologs are a central tenet of comparative studies including comparative genomics, phylogenetics, protein function annotation, genome rearrangement and structure. Using orthologs for comparisons are robust, and yield results that are applicable for one-to-one comparisons across evolutionary time scales. However, only focusing on orthologs limits understanding of novel genetic elements and does not explain the whole evolutionary history. Given that novel genetic elements are important to a lineage's genetic history and adaptability (Ohno, 1972; Force et al., 1999; Emes et al., 2003; Marques et al., 2008; Conant & Wolfe, 2008; Demuth & Hahn, 2009; Kaessmann et al., 2009; Innan & Kondrashov, 2010; Kaessmann, 2010; Rogers et al., 2010; Rogers & Hartl, 2012), ortholog-only comparisons will miss important variation that generates gene turnover and modifies transcriptome content and complexity over time. Complexity metrics from TranD are agnostic to orthology, as they present complexity independently from taxa being compared, using only within-species analysis on a single reference. Here, we compare the dynamics of conditioning on orthologs verses whole-transcriptome complexity metrics.

Annotations partitioned by ortholog have higher complexities across most metrics compared to the whole transcriptome (Fig. 2; Supp. Mat. Whole-Transcriptome Vs Ortholog Density Plots). In the Plantae group, for TpG, 16/44 taxa have ortholog complexities that lower than the whole-transcriptome complexity metrics and 2/44 for EpT. Fungi have 3/77 taxa where TpG complexity metrics are higher in whole-transcriptomes than in the ortholog set. Finally, 7/77 taxa across EpT and EpG in all the same taxa show the same complexity patterns above. The shift from higher complexity in ortholog datasets are due to exclusion of lineage-specific (non-orthologous) genes that are present in the whole transcriptome annotations. Novel gene complexities are substantially lower than genes with orthologs (Fig. 3; Supp. Mat. Novel Density Plots), consistent with prior work showing novel genes are less complex (Jacob, 1977; Khalturin et al., 2009; Siepel, 2009; Kaessmann, 2010; Jarosz et al., 2010; Tautz & Domazet-Lošo, 2011). This observation shows that many of the novel genetic elements are single-transcript and single-exon genes compared to that of the older orthologous genetic elements representing higher complexity transcripts. Whole transcriptome annotations contain both novel genetic elements and orthologous genetic elements. Hence the less complex novel elements drive differences in complexity compared with orthologs for all complexity metrics (EpT, EpG, TpG). Four well annotated species serve as examples for these trends (one species per clade). Distributions for *Homo sapiens*, *Drosophila melanogaster*, *Zea mays*, and *Neocallimastix californiae* all show significant differences, with p-values zero or near zero, between whole transcriptome and partitioned orthologous annotations (Fig. 2). Interestingly, novel genetic elements across all taxa shift to lower EpT compared to orthologous genes as most new genetic elements have lower transcriptome complexity than older orthologs (Fig. 3). An artifact observed in our data, between novel and orthologous datasets is the proportion of novel genes greater than 50% for *Gorilla gorilla* (western lowland gorilla) [GCF_008122165.1_Kamilah_GGO_v0], *Homo sapiens* [GCF_000001405.39_GRCh38.p13], and *Musa acuminata* (wild Malaysian banana) [GCF_000313855.2_ASM31385v2] (Supp. Fig. 2). See Supp. Mat. Whole-Transcriptome Vs Ortholog Density Plots and Supp. Mat. Novel Density Plots for a list of all organisms used in this study.

Across the Deuterostomia, mean TpG metrics between whole-transcriptome and ortholog annotations are significantly different in 66 of the 68 evaluated species. Only the Crested Ibis [GCF_000708225.1_ ASM70822v1] and the Florida lancelet [GCF_000003815.1_Version_2] have no significant differences for TpG metrics. When measuring complexity with mean EpT and mean EpG, all species showed significant differences between orthologs and whole-transcriptomes.

Across *Drosophila,* we observe a parallel pattern in complexity between ortholog and whole transcriptome data. Of the 11 species evaluated, 3 do not show significant differences for TpG under ortholog conditioning. These include *D. grimshawi* [dgri−all−r1.3], *D. persimilis* [dper−all−r1.3], and *D. sechellia* [dsec−all−r1.3]. These results could be an artifact of annotation performance and quality compared to more well-assembled species with molecular support for gene features, as many *Drosophila* lineages from the 12 Genomes Consortium were annotated by identifying ORFs not tuned to a specific lineage (Yang et al., 2018). Lineages that are not significantly different between whole-transcriptome and conditioning on orthologs vary in annotation completeness compared with other taxa, with a maximum TpG = 2. This is a noticeable departure from the other lineages that range from 45 maximum TpG *D. simulans* [dsim−all−r2.01] to 75 maximum TpG in *D. melanogaster* [dmel−all−r6.07]. In these instances, we suggest that the robustness of annotations varies between lineages and collapsed annotations combining exons from multiple transcripts have artificially skewed apparent transcriptome complexity (Shiao et al., 2015). In contrast, EpT and EpG metrics are all significantly different between whole-transcriptome and orthologs for all taxa.

In Plantae, across both EpT and EpG metrics, all samples (44/44 species) are significantly different when conditioning on orthologous. For TpG, 34/44 species show significant differences between whole-transcriptome complexity and complexity for orthologous sequences.

Only 1 fungus of 77 shows a significant difference between whole transcriptome and ortholog TpG, *Bactrachochytrium salamandrivorans* [GCA_002006685.1_Batr_sala_BS_V1]. All other fungi either do not deviate between whole transcriptome and ortholog TpG or only have TpG metrics that equal 1 for all genes in whole transcriptomes and subsequent ortholog partitions, i.e., no differences to compare. There are significant differences in EpT between orthologs and whole-transcriptome annotations for n = 53/77, with EpG having the same dynamics as EpT, where the same taxa (53/77) are significantly different between orthologs and whole transcriptome annotations.

Conditioning on orthologs yields complexity results that deviate from the whole transcriptome complexity. Of the 200 species used in this project 185/200 had significant differences under ortholog conditioning for TpG and 176/200 for EpT and EpG. These lower numbers are driven by fungi EpT and EpG having nearly identical mean EpT and EpG metrics. Across all lineages from each taxonomic group, a significant impact on complexity when orthology is required under all three metrics, except for mean TpG for fungi ($|T| = 47.333$, Fisher's combined P = 0.840, Supp. Table 1).

From our Jackknife cross-validation, 197/200 samples validate. Only 3 plant species differ with *Musa acuminata* (wild Malaysian banana) [GCF_000313855.2_ASM31385v2] not validating for

EpT and TpG not validating for *Pyrus x bretschneideri* (Chinese white pear) [GCF_000315295.1_Pbr_v1.0] and *Oryza sativa*, Japonica Group, (Japanese rice) [GCF_001433935.1_IRGSP-1.0].

Annotations can be facilitated by homology from better studied species, where the differences in metrics between orthologous partitions and whole transcriptome data are not significant. From the novel genetic element conditioning, we see that the variation of almost all species, is that ortholog metrics shift to higher complexities, while novel lineage specific genes have lower complexity. Hence, whole transcriptome data will show different values than either of these subsets. Future phylogenetic analysis must be aware that there will be biases from annotations, whether from highly rigorously annotated species or annotations from understudied organisms. Studies should consider how their data performs, and potentially correct for ortholog bias based on the questions and aims that are at hand.

*Relative Performance of Complexity Statistics*

Understanding how metrics perform in whole transcriptome data is essential to compare performance in evolutionary analysis of transcriptome complexity for analytical applications. If the number of isoforms depends on the number of exons, then we may expect correlations between EpT, EpG, and TpG. However, if isoform combinatorics decouple the number of exons and the number of isoforms, then we may expect distinct patterns from each metric as species diverge. In such a case, analysis with one metric could suggest evolutionary stability, masking real variation that could be apparent using alternative metrics. To diagnose these differences between our complexity metrics we estimated Pearson R correlation coefficients among each group.

All correlations are significant and positive (Fig. 4). TpG has the highest residuals from the regression line ($R^2$) when compared to the EpT and EpG metrics. While EpT and EpG metrics are more closely correlated. $R^2$ values are nearly one for both *Drosophila* ($R^2 = 0.962$, $P < 0.01$) and Plantae ($R^2 = 0.911$, $P < 0.01$). Deuterostomes show more moderate correlations for EpT vs EpG ($R^2 = 0.572$, $P < 0.01$), however, it is the highest $R^2$ value for the group. Deuterostomes have the largest variance between plotted metrics with TpG vs EpT ($R^2 = 0.370$, $P < 0.01$) and TpG vs EpG ($R^2 = 0.0964$, $P < 0.01$). *Drosophila* also show a significant correlation TpG vs EpT ($R^2 = 0.741$, $P < 0.01$) and TpG vs EpG ($R^2 = 0.831$, $P < 0.01$). Plants are moderately aligned closer to one in the $R^2$ values: TpG vs EpT ($R^2 = 0.603$, $P < 0.01$) and TpG vs EpG ($R^2 = 0.582$, $P < 0.01$). One non-vascular plant constituent, (Bryophta, *Physcomitrium patens*) the spreading earthmoss, is not clustered with our vascular plants, having the highest values for all metrics comparison within the group (See Supp. Fig. 3)

Most fungi annotations are single transcript genes and are therefore not amenable for similar analysis (Supp. Fig. 4). In current annotations, given that a significant majority of genes are single transcript, EpT and EpT metrics are highly correlated ($R^2 = 0.999$). Fungal genome sequencing and annotation is a burgeoning field and still has artifacts as genomic sequencing remains challenging (Wang et al., 2016) and annotation of isoforms in many species appears incomplete (Gordon et al., 2015) (Supp. Information, Supp. Fig. 5-16).

*Comparison to Broken Stick is Robust to Orthology*
Differences in complexity for orthologs and whole transcriptome analyses are driven largely by a bias in novel genes toward low complexity genes with few exons and few alternative transcripts. The Effective Exon Number (EEN) offers a metric that can compare the distribution of exon sizes within transcripts against random expectations of uniform exon placement, and accounts for differences in exon sizes (Hong et al. 2006). If EEN = EpT, splice junctions are evenly spaced, with all exons of equal size. Where EEN is far less than EpT, splice junctions are more uneven than expected under a uniform distribution, with over-dispersed exon sizes. While absolute patterns of EEN may be informative, comparisons of EEN to EpT characterize the deviation from a Broken Stick Model (Supp. Fig. 17) of randomly scattered intron positions. Across all species of chordates, *Drosophila*, and plants, EEN comparisons with the Broken Stick Model remain robust to conditioning on orthology and exclusion of lineage specific genes. We observe no difference in deviations from the Broken Stick Model when considering orthologs compared with whole transcriptome data. Only a minor shift upward is apparent in the lowest EpT values for Deuterostomes, *Drosophila,* Plants, and most Fungi (Supp. Fig. 18). Because the Broken Stick Model conditions on EpT, the effects of removing less complex lineage specific genes are largely mitigated so long as the remaining genes follow similar patterns of exon sizes.

Out of 68 deuterostome species, 62 have mean +/- 2SE EEN below the bound of the Broken Stick Model (Fig. 5). Humans, whose annotations are extensively validated and supported by abundant molecular evidence, lie among some of the lowest values suggesting more clustered intron breakpoints than the Broken Stick Model. These results suggest that molecular or evolutionary constraints on splicing processes or differences in annotation models are producing more tightly spaced intron breaks than one would expect based on random intron placement drawn from a uniform distribution. Human annotations are supported by exceptional molecular evidence and more well-developed curation efforts.

The remaining 6 species of chordates that show elevated EEN encompass diverse species that are not related phylogenetically or with respect to lab of origin (duck, turtle, lancelet, hedgehogs). Results do not shift significantly when conditioning on orthologs and excluding lineage specific genes. Hence, these metrics, unike Ept, EpG, and TpG, appear to be robust in the face of phylogenetic comparisons that require orthology. However, EEN conditions on the exon number and compares to expectations for randomly placed junctions. These are not sensitive to the subset of lineage specific genes that orthology excludes, which have few introns.

*Drosophila* show a different pattern. Transcripts with 7 exons or less show EEN greater than or equal to a Broken Stick model across all species. However, when transcripts have 12 exons or more, we begin to observe a departure with significantly lower EEN in in 5/11 *Drosophila* transcriptomes. Such metrics indicate more tightly clustered intron breaks than expected if intron breaks are randomly chosen from a uniform distribution. This pattern holds true even in the well validated model organism *Drosophila melanogaster.* Such results suggest that there may be different biochemical or evolutionary constraints on exon junction placement and associated splicing processes for genes with large EpT. There are 4/11 *Drosophila* that show elevated EEN, suggesting a more even distribution of intron breaks than expected from a random uniform distribution. Conversely, *D. sechellia*, only 0.25 years divergent from *D. simulans*, shows elevated EEN compared to the null model. *D. sechellia* annotations showed a lower density of transcripts with higher EpT values (Supp. Mat. Whole-Transcriptome Vs Ortholog Density Plots), and often are annotated with only one transcript per gene. Annotations that collapse isoforms with alternate exons into a single transcript will obscure the true distribution of EEN, biasing statistics toward higher values. Whole genome divergence in transcriptome complexity across such short timescales would be surprising. Whether the ultimate variation of EEN reflect biology or artifacts, these results suggest that cross-species comparisons of complexity require normalization for whole genome differences (Supp. Fig. 19).

The majority of plant transcriptomes lie above the Broken Stick model, with the highest EEN values in a club moss, *Selaginella moellendorffii*. Maize (*Zea mays*) aligns well with the Broken Stick model, but *Arabidopsis thaliana* lies well above. Sorghum agrees with maize (with overlapping error bars), except at the highest values of EpT (between 17-20 EpT) where it converges with *Arapbidopsis*.

Fungi show unusual variation in EEN compared with other clades. Most species show distributions of EEN centered about the Broken Stick model when EpT < 10. However, above 15 EpT, EEN increases toward more uniform spacing. These results appear more similar to patterns of exon distribution in plants than in animals. However, some fungi show atypical patterns as clear outliers in comparison with the Broken Stick model. A few unicellular fungi such as *S. cerevisiae* have few exons per transcript genome-wide, resulting in low variation in EEN, with little information via this metric. Fungi remain less well annotated in comparison with other clades, a gap that can now be addressed as cost and infrastructure for genome sequencing improves.

These results portray yet another class of variation among transcriptome complexity across taxa, which is fortunately robust to the effects of ortholog calling. Analysis of splicing patterns using EEN and comparison to models like the Broken Stick may be informative as annotation projects assess quality and compare molecular evolutionary variation (Supp. Fig. 20).

*Evolutionary Rate Analysis*
Understanding genetic complexity across the tree of life is essential to understand the processes that influence form and function of life. Analyzing genetic complexity in a phylogenetic context is important because biased estimation can lead to conceptually flawed interpretations of genetic function (Brown et al., 2001; Bininda-Emonds, 2004; Gevers et al., 2004). Furthermore, understanding the rate shift dynamics of complexities between orthologs and whole-transcriptomes gives empirical estimations to observe the variation between lineages' annotations containing novel genetic elements or shared genetic elements (orthologs). Here, we perform evolutionary analysis using transcriptome complexity as a genetic trait to identify shifts in complexity across taxa and to explore biases introduced in ortholog conditioning.

When estimating rates of complexity evolution across phylogenies we find that in some cases conditioning on orthologs causes a significant shift in rate estimates r, depending on the taxonomic group and metric being used. Here we used Phytools to perform posthoc tests of evolutionary rate changes between whole-transcriptome and orthologous rates. All metrics in the deuterostome group are significantly different: TpG (t = -3.89, P = 2.00E-04), EpT (t = -2.60, P = 0.0108), and EpG (t = 4.48, P = 0). We find no significant difference in the *Drosophila* group between complexity metrics: TpG (t = -0.564, P = 0.580), EpT (t = -0.123, P = 0.904), and EpG (t = -0.314, P = 0.7577). While in other groups we find some metrics are significantly different while others are not. Plantae orthologs compared to whole transcriptome metrics are as follows TpG (t = -2.18, P = 0.0335), EpT (t = 0.423, P = 0.667), and EpG (t = -1.68, P = 0.0972). Where mean TpG is significantly different between ortholog and whole transcriptome metrics, but non-significant for EpT and EpG. The Fungi group, among classes, had no significant differences between orthologs and whole-transcriptome for all metrics: TpG (t = -0.0161, P = 0.9872), EpT (t = -0.461, P = 0.648), and EpG (t = -0.463, P = 0.647).

By estimating credible rate shifts (each defined as an event) from the posterior probabilities (PP), we observe how rates differ between whole-transcriptome and ortholog complexities across a phylogeny for EpT (Fig. 6; See Supp. Fig. 21-24 for MCMC convergence). In deuterostomes we identify a shift (e.g., warmer heat color) for taxa related to humans. At the simian node [from *Callithrix juccus* (common marmoset) to *Homo spaiens*] in the phylogeny containing whole-transcriptome EpT, we observe an elevated evolutionary rate [PP = 0.0972]. While the ortholog EpT evolutionary rate is also elevated at crown primates [PP = 0.149] There is also an elevated rate for the branch of *Branchiostoma floridae*. The rate is lower in whole-transcriptome EpT [PP = 0.0679] but higher, given the edge taxa, in the orthologous dataset [PP = 0.108]. Outside of *Branchiostoma floridae*, the rates on the phylogeny illustrates that all chordates have similar rates outside of primates. Some of these rate shifts may be due to the artifact of human and primate genomes having more complete annotation with greater molecular evidence to support isoform detection.

*Drosophila* have an elevated rate of EpT at the node between *D. sechellia* and *D. simulans* for both whole-transcriptome [PP = 0.643] and orthologs [PP = 0.649], consistent with the *D. sechellia* annotation containing transcript models with a union of exons from multiple transcripts within a gene (Shiao et al., 2015). Elevated rate shifts within plants and deuterostomes occur at single branches of organisms that are highly annotated. Phylogenetic branch tips that have high-rate shifts include *Arabadopsis thalana* and *Camelina sativa* (false flax) [PP = 0.0592] and *Quercus suber* (cork oak) [PP = 0.110]. These plants are highly studied cash crops or, in the case of *Arabadopsis*, a genetic model system. While orthologous genes have a rate shift in *Zea mays* (maize) [PP = 0.278] with a higher mean EpT than other Potales, no shift is observed in whole-transcriptome annotations. There is a rate shift at the basal portion of the tree between vascular and non-vascular plants. This rate shift is lower in ortholog genes (PP = 0.00852), than it is whole-transcriptomes (PP = 0.0162). With larger genome size, plants contain more genes and a plethora of extra functions evolved in vascular plants compared to that of mosses (Bainard et al., 2020). Hence, we suggest these differences are likely biological. The ploidy levels of plants vary tremendously among the group, and annotation comparisons may not directly be comparable

There are rate shifts within fungi lineages, including a split between Chytridiomycota and all other fungi phyla (Basidiomycota and Ascomycota), PP = 0.0301 in whole-transcriptomes and PP = 0.0427 in orthologs. Another shift is observed in the class branch Pneumocystidomycetes, PP = 0.396 in whole-transcriptomes and PP = 0.570 in orthologs. The rate shifts are higher in orthologous genes than in whole-transcriptomic genes.

Biologically, complexity rate shifts are relatively rare across the tree of life (Fig. 6). We only observe high shifts in deep time or shift among highly studied biological systems. Shifts in complexity among branches may stem from annotation quality or vigor, given the specific taxa where shifts occur. The difference between whole-transcriptomes and orthologs have higher probabilities in rate shifts among orthologous dataset than of whole transcriptomes. Overall, these observations suggest that analysis of evolutionary rates is not severely impacted by ortholog conditioning, unlike species-level complexity analyses.

## Discussion:
**Analysis of Transcriptome Complexity**
Over 95% of multi-exon genes are subject to alternative splicing in eukaryotes (Pan et al., 2008; Wang et al., 2008). As the genomic field expands, we have found that a tremendous amount of complexity is not found simply found in how many single genes a genome possesses, but instead of driven by genetic machinery "cutting" and arranging genetic units for specific tasks. These dynamics are observed in the human genome with only ~25k genes coding upwards of ~90k proteins (Venter et al., 2001; Valdivia, 2007). These genetic mechanisms and genomic dynamics add extra layers to understand organismal complexity. Understanding complexity is a central tenet to evolutionary biology. As mutations accumulate, novel genes form and gain function speciation events occur, yielding new species. Understanding the processes that drive complexity across the tree of life, inherently, has the potential to illuminate biological diversity that we observe in nature.

Transcriptome complexity is influenced by many biological factors. Previous work has observed exceptional splicing patterns in the testes and heads of *Drosophila*, mice, and humans (Reviewed in Barbosa-Morais et al., 2012; Merkin et al., 2012; Gibilisco et al., 2016; Naro et al., 2021). Alternative splicing in males and females produces functional differences in sex determination pathways (Reviewed in Salz, 2011). The use of different transcripts across timepoints and tissues influences animal development and complexity of body forms (Gustincich et al., 2006). Changes in alternative splicing and the addition of complex combinations in isoforms can allow for greater functional diversity and drive evolutionary innovation (Gilbert, 1978). Hence, a clear need exists for metrics that elucidate whether and how evolution has reshaped transcriptome complexity in different organisms. Here, we use new metrics generated by TranD to measure trends of complexity accurately and precisely across taxa. Understanding complexity is quite important to understanding the dynamics that facilitate the machinery of life (McShea & Brandon, 2010; Day, 2012).

The metrics presented here are robust since they are agnostic in ascertainment given the data. These metrics also allow for independent diagnosis of complexity for a single organism along with the ability to compare each metric. Caveats for use cases are largely driven by the quality of annotation input data. TranD will collapse and collate transcriptome complexity across genes only as accurately as the annotation data at hand. We observe such effects in a few constituents, such as in Deuterostomia, the primate group has higher than average metrics. This is probably induced by the human genome annotation. Being one of the most studied genomes, it has a wide breadth of data associated with it. This allows for closely related taxa to, in turn, use that resource for more complete annotation than more distantly related organisms. In contrast,

*Drosophila sechellia, B. floridae* and whale sharks may show signs of collapsed isoforms among annotations, which yield unusual complexity results (Supplementary Information).

**Ortholog-free complexity comparisons capture genetic novelty**

Conditioning on orthologous genetic elements for complexity metrics, in our findings, induces bias for more complexity in older, more studied genetic elements, not taking into account the novel gene formations that is only found in whole-transcriptome annotations. This bias will have complexity metrics that capture more AS schemes, in turn, being more complex from the whole-transcriptome complexities that take into account all the novel elements. These novel genetic elements arise from either re-organization of pre-existing genes or are *de novo* (Kaessmann, 2010; Tautz & Domazet-Lošo, 2011). Since novel genetic elements are, indeed, "new" less time has allowed for genes to accumulate more AS schema yielding more isoforms, hence more complexity in this study's metrics. Using orthologous genes in phylogenetic comparisons allows for one-to-one comparisons given a speciation event (Rentzsch & Orengo, 2009; Studer & Robinson-Rechavi, 2009) dubbed the "ortholog conjecture" (Nehrt et al., 2011). This methodology is a "necessary evil" in the genomic age, because much of the data generated may not adhere to all data being of orthologous origins. Even in relatively recent studies, across multiple taxa using many genomes, have lacked the resolution for concordance of a single gene copy (Thomas et al., 2020). This is also not a unique occurrence in whole-genome phylogenetic studies as more taxa are incorporated (Emms & Kelly, 2018; Smith & Hahn, 2021).

More studies have tested the notion if discarding paralogs is indeed the best practice (Studer & Robinson-Rechavi, 2009; Stamboulian et al., 2020; Smith & Hahn, 2021). Stamboulian et al. (2020) observed that paralogs still aid in functional prediction and throwing away huge swaths of data can often lead to the lack of predictions because no orthologs are available. Recent work has demonstrated that paralogous data should not be as feared as once thought for phylogenomic inference, reviewed in Smith & Hahn (2021); (De Oliveira Martins et al., 2016). Of course, this paper uses EEN to normalize complexity discrepancy between whole-transcriptome and orthologous complexity metrics. Using Hong et al.'s (2006) effective exon lengths model, EEN allows for whole-transcriptome and orthologous metrics to me normalized by the same scale. Thus, making a more concordant comparison between various datasets without the need to discard non-orthologous data (See Results: *Comparison to Broken Stick is Robust to Orthology*). Even with corrections to mitigate biases and loss of potentially informative data, we still recommend researchers use best practices for their experimental designs be it include whole-transcriptomes, condition on orthologs, or incorporate the EEN correction.

Novel lineage specific genes frequently appear with fewer exons and lower protein complexity than background genomic properties (Wang et al., 2005). As new genes are added to the transcriptome, these sources of innovation may therefore add sequences that have unusual complexity compared with long-standing sequences. Our analysis confirms this hypothesis, and these effects are observed in comparisons of orthologs and ortholog-free whole transcriptome analysis. We find that novel genes have lower complexity with respect to every metric for analysis. If complexity analysis only uses orthologous sequences, the unique properties of these novel sequences will be obscured. Moreover, bias in complexity may offer a false portrait of whole transcriptome complexity in different organisms. In some cases, these biases may be so severe that evolutionary rate analysis is skewed to false significance, as we observe in primates. However, in other use cases, such as comparisons of EEN to Broken Stick, novel genes do not alter results significantly as they condition on baseline complexity metrics of EpT. In future applications, users will undoubtedly wish to assess the impact of ortholog conditioning on the questions at hand, potentially with analytical corrections for biases. With quantitative, precise descriptions of complexity biases for orthologs and new genes, we can move forward with evolutionary analysis with greater power than if accepting orthologs solely as a "necessary evil". The working examples from analyses presented here show how transcriptome analysis can be implemented for future insights as the number of successfully sequenced and annotated organisms expands, especially in previously non-model systems.

**Metrics and use cases for future evolutionary analysis**
This study offers examples for how metrics of transcriptome complexity may reveal differences in genomes for divergent taxa. These complexity metrics can be utilized further than this current study to analyze newly annotated genomes in emerging systems. With the appearance of well annotated transcriptome datasets, it may soon be possible to determine how biological factors influence isoform variation and alternative splicing patterns. TranD is not only limited to publicly available annotations but can calculate complexity metrics on *de novo* annotations with a pipeline of choice like BRAKER or a BUSCO analysis (Hoff et al., 2019; Manni et al., 2021). It is, however, advised to pay close attention to annotation quality and completeness when using TranD to estimate complexity metrics, since the results are highly sensitive to the data given. We have observed markedly different modes of transcriptome complexity among and between lineages across the tree of life.

From high complexity in model systems compared to their sister taxa as seen in *Zea mays* and humans to the simpler transcriptome forms identified in many fungi, *Arabidopsis*, and *Drosophila*. Therefore, potentials for biases in annotation of highly studied organisms compared to those less studied. Understanding the molecular and cellular underpinnings for such variation may reveal insights into fundamental biology in future studies. Similarly, new analyses similar to our current study may help us characterize observable phenomena in transcriptome evolution, with potential to illuminate processes, mechanisms, and dynamics of evolution in the tree of life.

As field of biology continue to progress and generate more precise and accurate genomic sequences, these metrics will hold even more power to understand the role complexity on the tree of life. These metrics are broad in nature, being able to be applied to any organism that possess a genome. For example, multicellular organisms are assumed to be more complex than unicellular representatives. Multicellular organisms have larger sizes and tissue types likely a caused by novel adaptation giving rise to new functional outlets for the transcriptome to evolve (Levine & Tjian, 2003; Sealfon et al., 2021). Our metrics can be used to guide further research to understand the phenomenon of more lineages with less complex body plans and traits have more complexity in through genomes. As future genomic resources emerge, it may be possible to compare single celled and multicellular relatives to infer how transcriptomes evolve as novel body forms emerge. Similarly, we may ask whether genomic complexity correlates with transcriptome complexity, or how sexual reproduction compared to asexuality influences complexity fluctuations. Understanding complexity systematically can help to expound patterns in evolutionary tracts across the tree of life where different dynamics have a role in evolution.

*Conclusion*
This study illustrates the power and utility using TranD transcriptome complexity metrics as both a comparative method and to independently validate complexity across any and all lineage of life and viruses given an annotation file. This work is a first step to elucidate complexity patterns and validate using whole-transcriptome sequences versus conditioning on orthologous genetic elements across the tree of life. Further research should be conducted on complexity to understand the tempo of evolution. Co-evolutionary patterns between host and parasite genomes, tempo of evolution in specific groups, wild-type vs domesticated genome complexity, and unicellular verses multicellular dynamics are all interesting questions that can be addressed using these metrics of transcriptome complexity. Furthermore, the ability to validate annotation files for specific studies has utility downstream. The creativity of the scientific question at hand and the annotation available are the only bottlenecks when using these agnostic transcriptome complexity metrics.

## Acknowledgements:
We would like to thank Dr. Rhyker Ranallo-Benavidez and Dr. Farnaz Fouladi for their help and knowledge providing insights into code and code accuracy. Furthermore, conversations with both Dr. Denis Machado and Dr. Alex Dornburg helped to provide proper statistical analysis for phylogenetic methods. Their help is much appreciated.

## Funding:
This study was funding with NIH NIGMS R35GM133376 (RLR) and GM128193 (LMM).

## Data Acquisition and Code:
All data, links to data used, and code can be found at
https://github.com/jemcquillan/Transcriptome_Complexity and
https://github.com/jemcquillan/OrthoDB_Parser.

**Glossary of Complexity Terms**

Alternative Splicing (AS): Molecular mechanism that modifies pre-mRNA constructs prior to translation, which produces a diverse set of mRNA from a single gene.

Transcripts per Gene (TpG): The number of transcripts that can be constructed from a single gene that have been annotated.

Exons per Transcript (EpT): The exons that are annotated within single transcript produced from a single gene.

Exons per Gene (EpG): The union of unique exons annotated within a single gene.

Effective Exon Number (EEN): A distribution of exons across a complexity metric (or gene), given the relative length of the exon to the total length of the region ($L_e$).

Ortholog: One of a set of homologous genes that have diverged from an ancestor as a consequence of speciation.

Novel Gene: Genes that have emerged inside a defined time frame are novel genes. Novel genes are will be classified by their age. The time frame is not fixed and needs to be defined for each study.

Evolutionary Rate: As defined through Rabosky et al., 2014 interpretation. A macroevolutionary rate dynamic that applies to some part of a phylogenetic tree. All lineages that share a common regime have exactly the same macroevolutionary rate at a given point in time

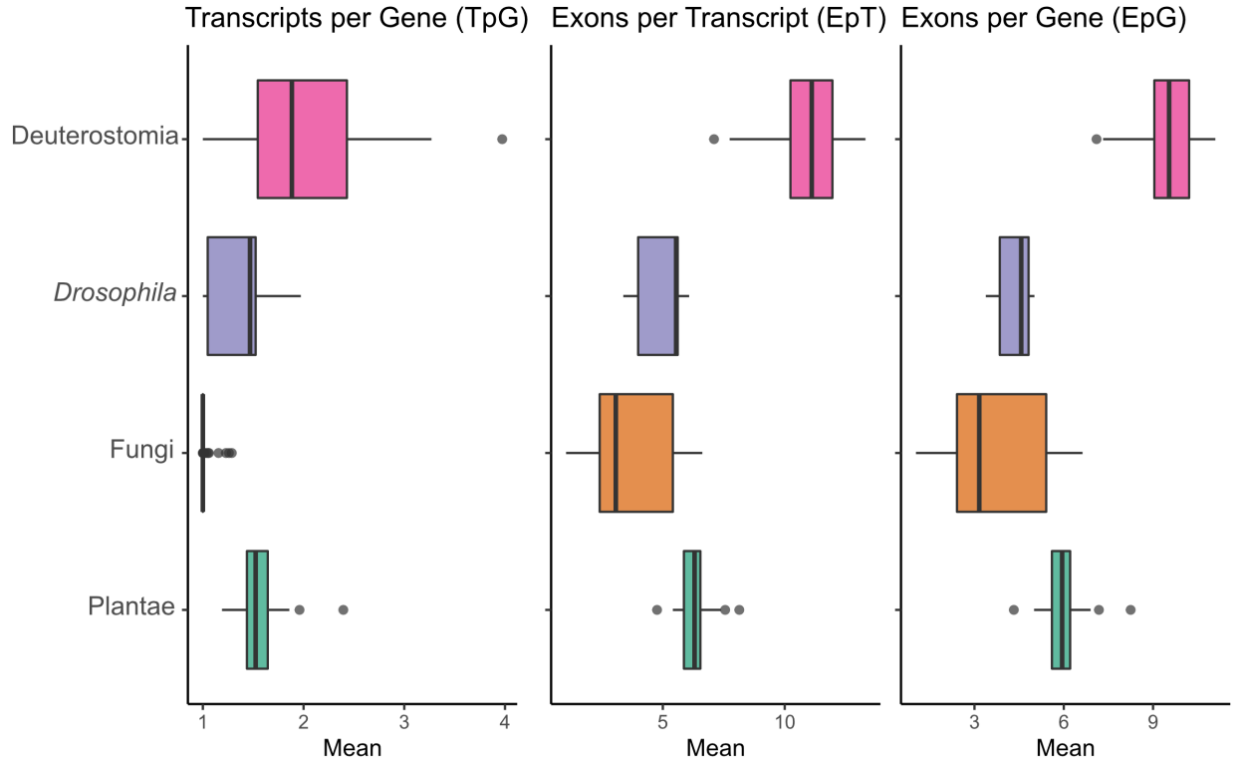

Figure 1: Box and whisker plots showing complexity metrics for TpG, EpT, and EpG for broad taxonomic groupings. Black lines indicate median metrics with the first and third quartiles at the borders of the box. The minimum and maximum range indicated by black lines. Phylogenies, on the left, were generated by TimeTree. Columns are set to each complexity metric. Taxonomic groups consist of A) Deuterostomia, B) *Drosophila*, C) Plantae, and D) Fungi.

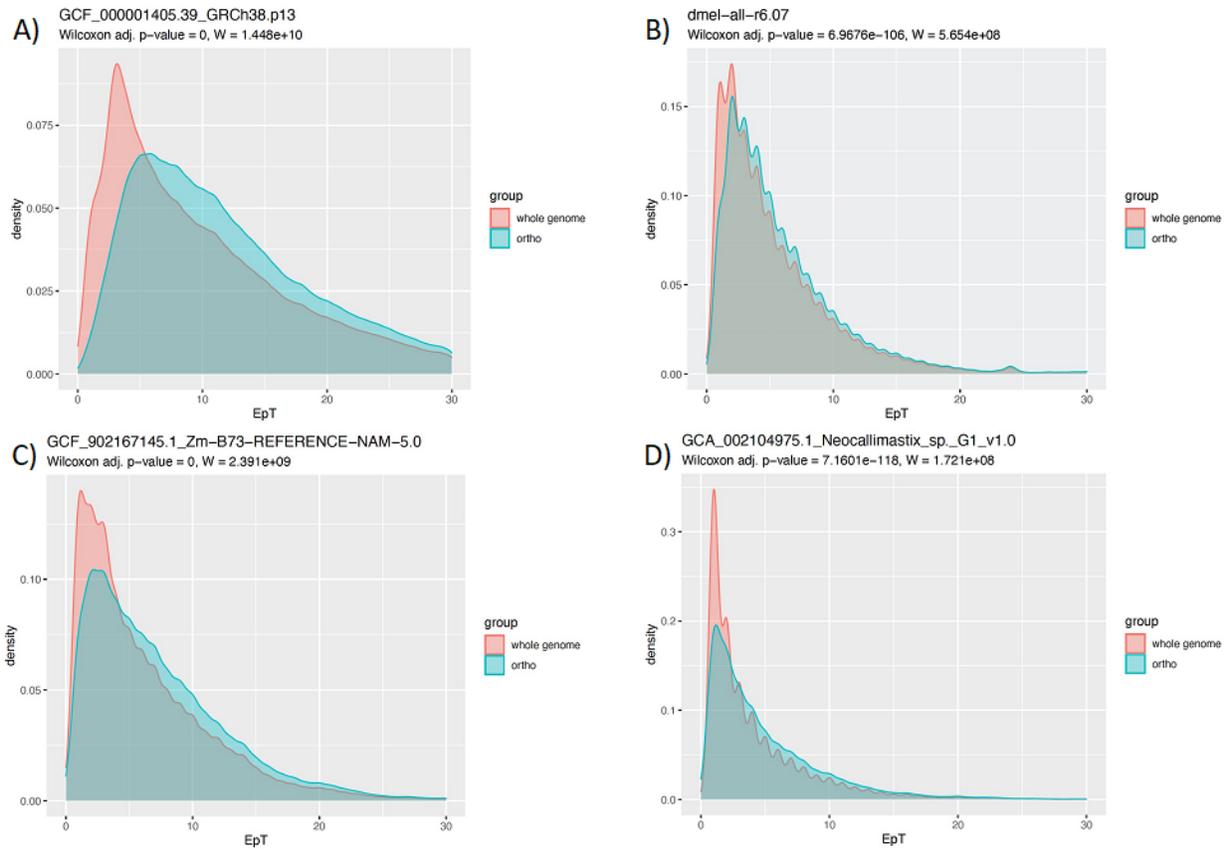

Figure 2: Density plots between whole genome (red) and orthologs (light blue) for Exons per Transcript (EpT) complexity metrics with densities on the y-axis and EpT on x-axis. Panels are truncated to 30 exons for purposes of visualization. We observe a significant difference in EpT for whole transcriptome annotations and partitioned by ortholog annotations in each of 4 example species using Wilcoxon rank sum test. A) *Homo sapiens*, W = 1.448 x $10^{10}$, P = 0, B) *Drosophila melanogaster*, W = 5.654 x $10^9$, P = 6.9676 x $10^{-106}$, C) *Zea maize*, W = 2.391x$10^9$, P = 0, and D) *Neocallimastix californiae*, W= 1.721 x $10^{10}$, P = 7.166 x $10^{-118}$. Across all taxa, ortholog conditioning results in data that are less likely to include low complexity transcripts.

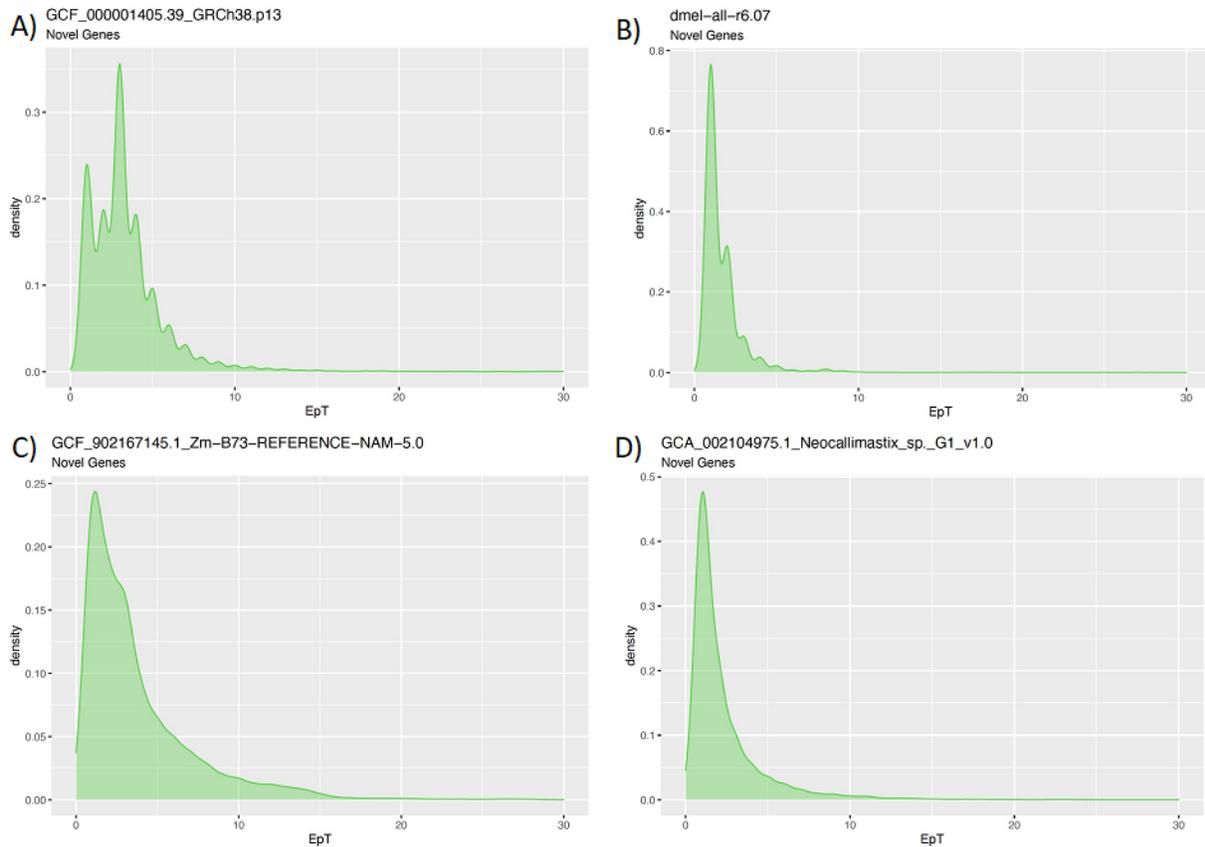

Figure 3: Density plots of EpT complexity metrics for novel lineage specific genes with densities on the y-axis and number of exons plotted on x-axis. Panels are truncated to 30 exons for visualization. Each panel A) *Homo sapiens* show a shift in the number of exons where most new genes have fewer than 10 exons per transcript with a mean EpT of 3.431. B) *Drosophila melanogaster* concentrates most novel genes between 1 (~61.2%) and 2 (~24.7%) exons per transcript with a mean EpT of 1.863. EpT has few transcripts with complexity higher than 2 exons per transcript. C) *Zea mays* follows a similar pattern with a shift left to lower numbers of exons per transcripts with a mean EpT at 3.822. Only ~27.7% of novel genes in maize are 5 EpT and higher, while under 5 is ~72.3% of all novel genes. Finally, D) *Neocallimastix californiae* novel genes show exon per transcript counts have a mean EpT of 2.321. Most of the distribution is concentrated in below 5 EpT (~88.4%). Values are substantially lower than observed for Orthologs and Whole Transcriptome.

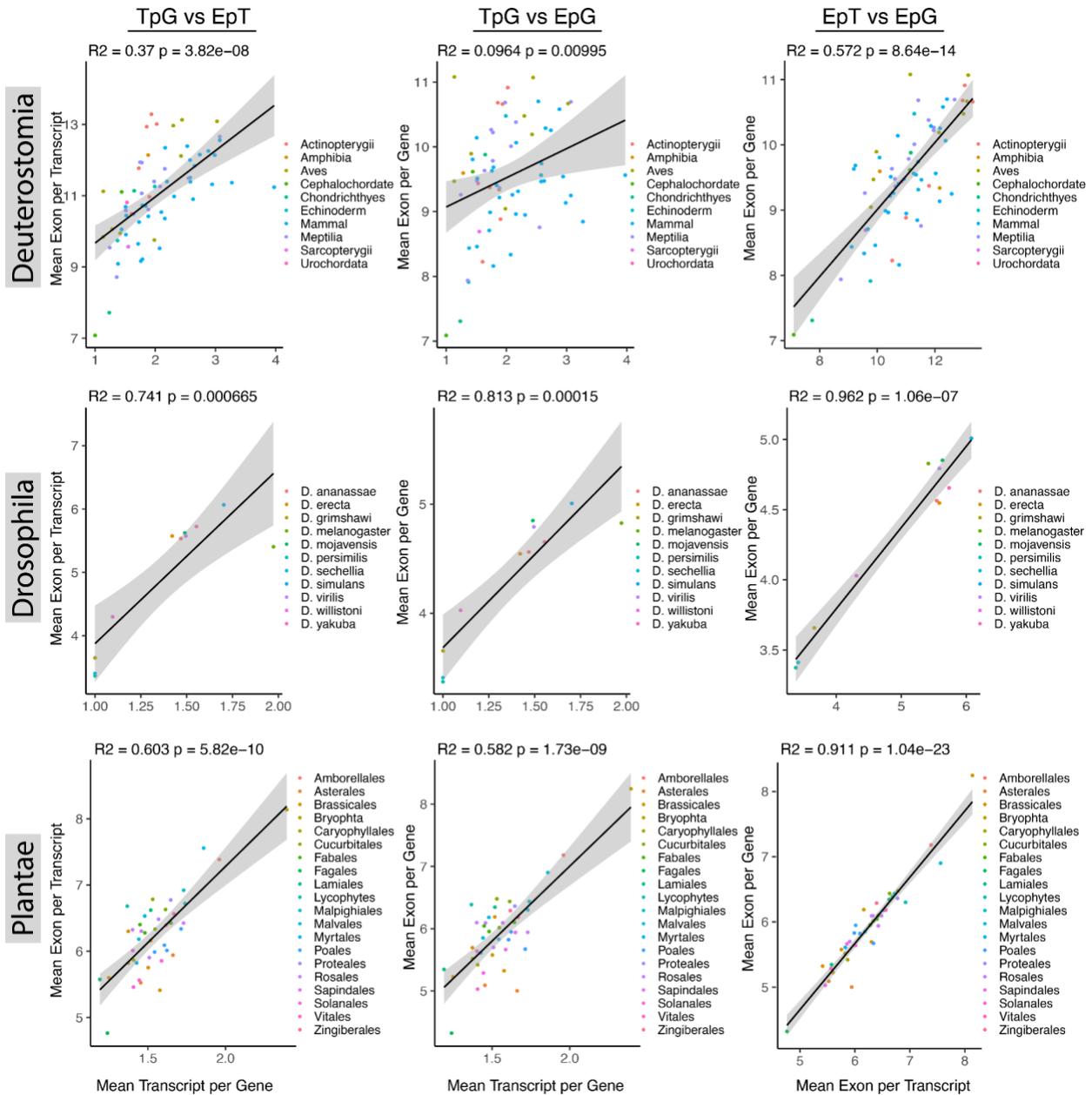

Figure 4: Pair plots comparing complexity metrics. Metrics are compared among TpG, EpT, and EpG. The p-value and $R^2$ derived from calculating the Pearson correlation coefficient. Taxonomic groups consist of Deuterostomes, *Drosophila*, and Plantae.

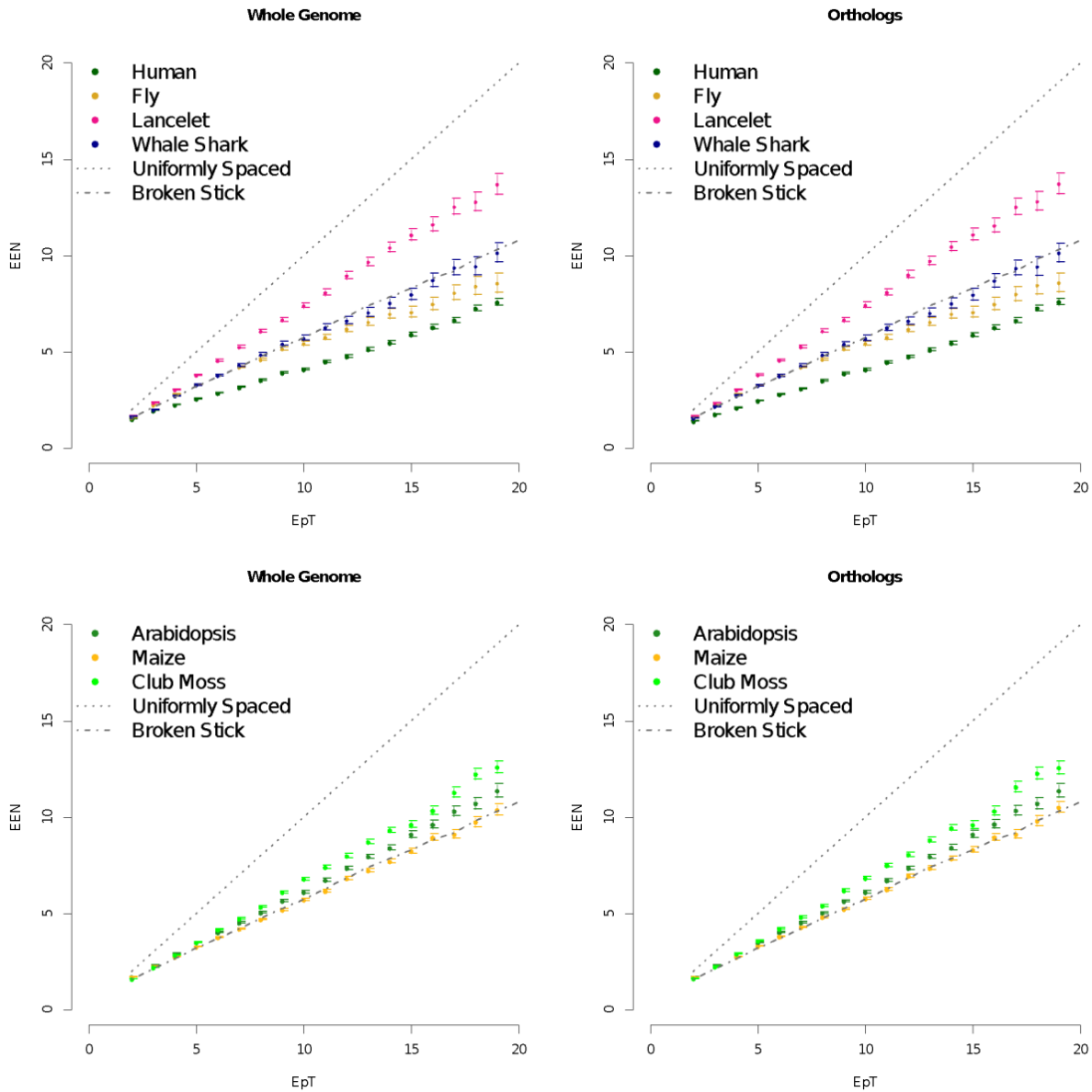

Figure 5: Effective Exon Number (EEN) vs Exons per Transcript (EpT) (mean EEN +/- 2*SE) in 4 species of animals and three species of plants. A/C) Whole Transcriptome data and B/D) after conditioning on Orthologs being present in at least one species in the phylogeny. Ortholog data excludes lineage specific genes. EEN is expected to follow a Broken Stick model if intron bounds are randomly drawn from a uniform distribution. Humans and *D. melanogaster* show lower EEN with a bigger effect at transcripts with high EpT values. Whale sharks show EEN fully consistent with the null expectation. Lacelets show elevated EEN, the highest of any Deuterostome, suggesting more evenly distributed exon sizes than random. While orthologs show a nominal shift in the smallest EpT values, requiring orthology does not alter comparisons to a broken stick model. Hence, Broken Stick Model comparisons are likely to be robust to effects of orthologs and exclusion of lineage specific genes in evolutionary analysis.

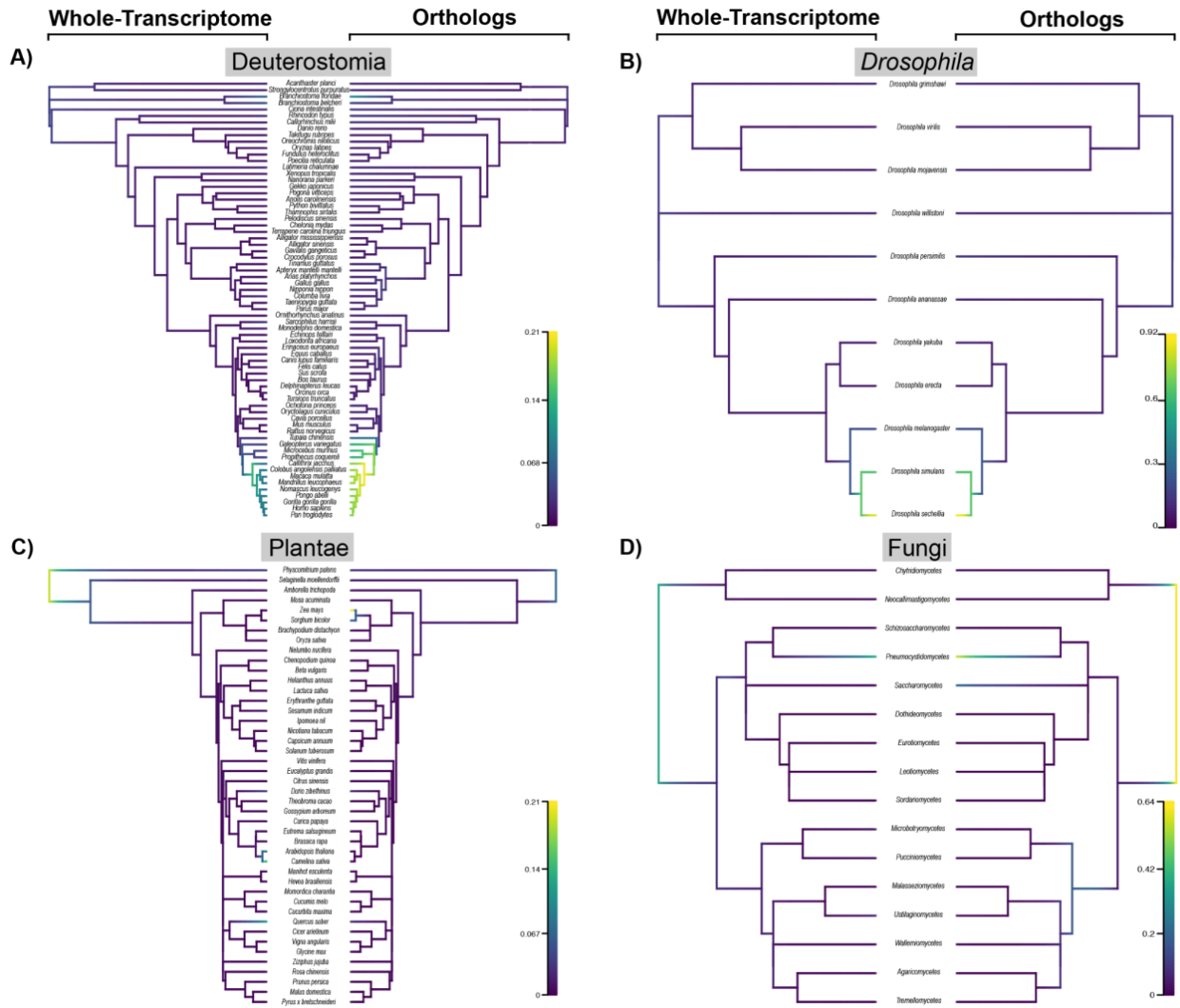

Figure 6: Evolutionary rates of EpT complexities across phylogenies for A) Deuterostomia, B) Drosophila, C) Plantae, and D) Fungi. Fungi phylogenies are collapsed to classes. Phylogenies were generated from the TimeTree Portal. Rates were calculated in BAMM with 10,000,000 generations with a 10% burn-in. Plots were generated in BAMMTools R package. Evolutionary rates compare whole transcriptomes (left trees) to orthologous genes (right trees). We observe a significant difference in evolutionary rates for Deuterostomia when conditioning on orthology, but not in other clades. Phylogenetic trees are scaled within each clade, with yellow being higher and blue being lower rates.

# SUPPLEMENTARY INFORMATION

## COMPLEXITY DIFFERENCES ACROSS GROUPS

Among all the higher lineages in this study, clades within have many constituents and a variety of metrics within a group. Deuterostomia classes have a broad range of values for each metric (Fig. 1). Observing TpG, ray-finned fish (Actinopterygii) are relatively conserved, with a range across mean TpG from 1.60-2.02 (a difference of 0.417) while Aves has a broad range, respectively compared to other deuterostomes, with a difference of 1.04 mean TpG and a range from 1.98-3.02. Mammals have a relatively high spread across the quantiles compared to other deuterostome classes (difference of 1.04) with outlier *H. sapiens* at 3.97 mean TpG, being the highest TpG complexity metric across every taxon in this study. Among all deuterostomes mean TpG is not significantly different among all deuterostome classes (KW $Chi^2$ = 15.35, P = 0.082). Mean EpT is also not significantly different among classes of deuterostomes (KW $Chi^2$ = 9.50, P = 0.393). Two taxa pull the width of the plots down in Cephalochardata and Chondroichthes the Florida lancelet (*Branchiostoma floridae*) at 7.10 mean EpT and the whale shark (*Rhincodon typus*) at 7.74 mean EpT. Certain dynamics change across EpT to EpG among different deuterstome classes. Amphibia become narrower from EpT to EpG, while Actinopterygii, Aves, Chondroichthes, and Echinodermata become wider. In total, EpG have lower complexity metrics in EpG than EpT. Among the deuterostome classes EpG are also not significantly different (KW $Chi^2$ = 11.02, P = 0.274). Complexity metrics can also be used to understand annotation quality between organisms, as observed in the stark differences between *Branchiostoma flordae* vs *B. belcheri*.

Drosophila complexity metrics between species are all significantly different TpG (t = 14.26, P = 5.82e-08), EpT (V = 66, P < 0.05), and EpG (V = 66, P < 0.05). The most robust annotations within Drosophila are from *D. melanogaster*. Like other model systems, it has the highest mean TpG, however, *D. simulans* has the highest mean EpT and EpG. Curiously, *D. ananassae*, *D. erecta*, *D. mojavensis*, *D. virilis*, and *D. yakuba* along with *D. simulans* all have higher a higher mean EpT complexity, while *D. mojavensis* and *D. simulans* have a higher mean EpG complexity value than *D. melanogaster*.

Among plantae orders mean complexity metrics are not significantly different from each other (mean TpG: KW $Chi^2$ = 27.47, P = 0.094; EpT: KW $Chi^2$ = 33.34, P = 0.022; EpG: KW $Chi^2$ = 33.65, P = 0.020). There is an outlier among all Plantae orders, from the bryophyte species *Physcomitrium patens*, spreading earthmoss. *P. patens* has a mean TpG 2.39, mean EpT 8.14, and mean EpG 8.25. These are much higher than the vascular plant clade in Viriplantae. Viriplantae complexity metrics have IRQR ranges between 1.43-1.65, 5.86-6.56, and 5.59-6.26 for mean TpG, EpT, and EpG, respectively. *P. patens* is outside of all complexity metric IRQ values in the group.

Fungi complexities are widely variable between classes among annotated taxa. The fungus lineages are old and highly diverse, as well a relatively unknown broadly. Thre is not a significant difference between mean TpG metrics across classes, KW $Chi^2$ = 17.84, P = 0.399. There are only a few outliers in this metric, with Chytridiomyces having the highest width. However, across all taxonomic classes the range is narrow being concordant at mean TpG 1.00 with only outliers being above 1.25 mean TpG. Both EpT and EpG mean complexity metrics are nearly identical. However, among the groups within Fungi there is significant difference for each metric among Fungi classes, EpT: KW $Chi^2$ = 66.49, P = 8.58e-08 and EpG: KW $Chi^2$ = 66.44, P = 8.76e-08. Patterns are easier to observe by consolidating Fungi classes into their respective phylum. EpT and EpG mean complexity is clearer to see that Ascomycota is lower and has nonoverlapping IRQ values than Basidiomycota, Chytridiomycota, and Mucoromycota. When Ascomycota is included among all phyla, there is a significant difference among taxonomic groups for mean EpT (KW $Chi^2$ = 40.49, P = 8.39e-09) and mean EpG (KW $Chi^2$ = 40.47, P = 8.47e-09). Without including Ascomycota in the KW-Test, there is not significant difference among the other Fungi phyla for mean EpT (KW $Chi^2$ = 5.151, P = 0.076) or mean EpG (KW $Chi^2$ = 5.151, P = 0.076).

Supp. Table 1: Fisher Combined Probability Test p-values and T statistic for each taxonomic group's complexity metric used in this study.

|  | TpG | | EpT | | EpG | |
|---|---|---|---|---|---|---|
|  | $|T|$ | P | $|T|$ | P | $|T|$ | P |
| **Deuterostomes** | $10^{-16}$ | $10^{-16}$ | $10^{-16}$ | $10^{-16}$ | $10^{-16}$ | $10^{-16}$ |
| *Drosophila* | 397.765 | $10^{-16}$ | 2637.79 | $10^{-16}$ | 2332.525 | $10^{-16}$ |
| **Plantae** | 3606.728 | $10^{-16}$ | $10^{-16}$ | $10^{-16}$ | $10^{-16}$ | $10^{-16}$ |
| **Fungi** | 47.333 | 0.8403028 | 10178.283 | 0 | 10165.276 | 0 |

ABNORMAL COMPLEXITY AND ANNOTATION
There are also potential signs of *B. floridae* having less resolution in its annotation compared to other deuterostomes, given its lack of complexity. Especially when comparing its sister taxon *B. belcheri*. Since formulating this manuscript, *B. floridae* has a new genome assembly (RefSeq assembly accession: GCF_000003815.2) but is not updated on OrthoDB. The previous assembly is from 2009 while the new assembly from 2020 is able to utilize the new genomic resources in both sequencing and *in silco* techniques. The new annotation has mean complexity metrics more in line with its placement in the deuterostome phylogeny. Receptively, *B. floridae* GCF_000003815.2 has mean metrics TpG: 1.58, EpT: 10.98, and EpG: 8.90; while *B. floridae* GCF_000003815.1 has mean metrics TpG: 1.00, EpT: 7.10, and EpG: 7.10. These metrics further explain the caveat to the metrics are sensitive to the annotation data being supplied.

When plotting complexity metrics across the deuterostome phylogeny, we observed a pattern where high outlier counts for a specific gene were missing in *Gallus gallus*, chicken. Upon further investigation, the gene was identified as Titin. Titin is the largest known described protein that has a significant role in structural, developmental, mechanical, and regulatory roles in cardiac and skeletal muscles (Fürst et al. 1988; Vikhlyantsev et al. 2012). We found that the previous *Gallus gallus* annotation, build 6a (RefSeq assembly accession: GCF_000002315.6)

(Bellott et al. 2017), did not have Titin annotated while the current annotation (RefSeq assembly accession: GCF_016699485.2), used in this paper, does have an annotated *TTN* gene has annotated Titin. This is a key example of how TranD can be used to guide and help utilize annotations properly to obtain the most accurate and complete dataset, especially as annotation and assembly progresses rapidly in the face of new data. This result illustrates the power of complexity metrics in identifying patterns in structural phenotypes of transcripts along with TranD being dependent on proper annotations for accurate metrics.

The fruit fly, *D. sechellia*, annotation has lower complexity metrics compared to its other congeners. The *D. sechellia* annotation collapsed many of its transcripts and exons into genes, thus having metrics that are mostly single transcript genes, without much AS, compared to the rest of the *Drosophila* clade. The Drosophila 12 Genomes Consortium; et al. (2007) found there to be "appreciably higher frequency of masked bases in lower-quality" in the *D. sechellia* assembly and it is observed in our complexity analyses.

# TAXONMICS SCALE RESULOUTION COMPLEXITY METRICS

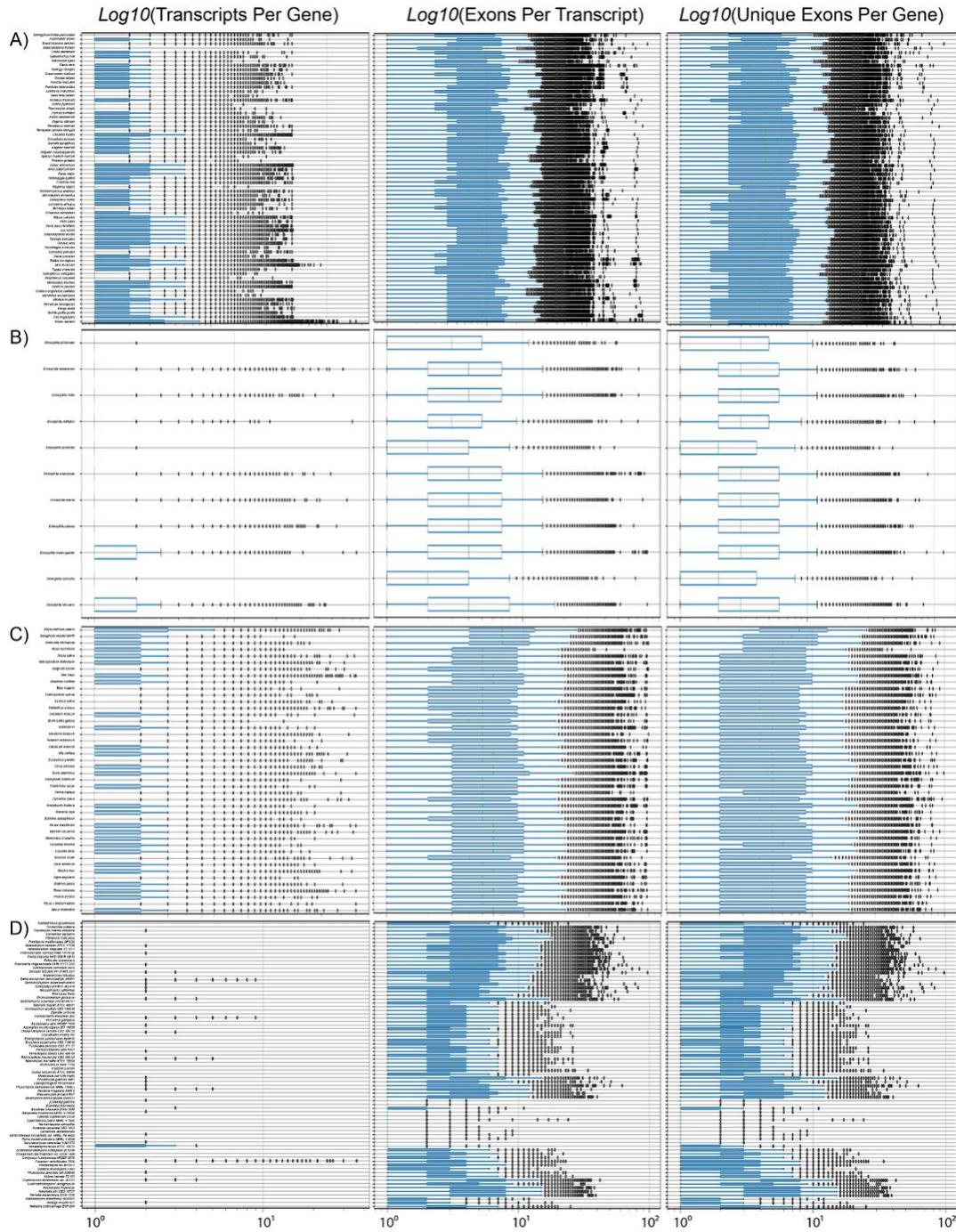

Supp. Figure 1: Box and whisker plots showing complexity metrics for TpG, EpT, and EpG for individual taxonomic groupings in log-scale. Green lines indicate median metrics with the first and third quartiles at the borders of the box. The minimum and maximum range indicated by black lines. Columns are set to each complexity metric. Taxonomic groups consist of A) Deuterostomia, B) *Drosophila*, C) Plantae, and D) Fungi.

**NOVEL GENE PROPORTIONS**

Supp. Figure 2: Proportions of novel genes (Red) to orthogous genes (aqua) for each oraganisms annotations. Panels are broken down by taxonomic groups consist of A) Deuterostomia, B) *Drosophila*, C) Plantae, and D) Fungi.

## VARIANCE

Of all our metrics are primarily over-dispersed compared to expectations under the Poisson (Supp. Fig. 2) Both EpT and EpG for all deuterostomes, *Drosophila*, and plants are over-dispersed compared to expectations under the Poisson Distribution. Fungi have 59/77 taxa over dispersed EpT metrics and 60/77 for EpG. All fungi taxa are under-dispersed for TpG. TpG is have more taxa under-dispersed compared to expectations under the Poisson Distribution than other the other complexity metrics. TpG is over-dispersed for 57/68 Deuterostomes, 7/11 Drosophila, and 33/44 plants taxa. We observe exceptionally high normalized variance in humans across TpG and EpT with variance two orders of magnitude higher than the mean.

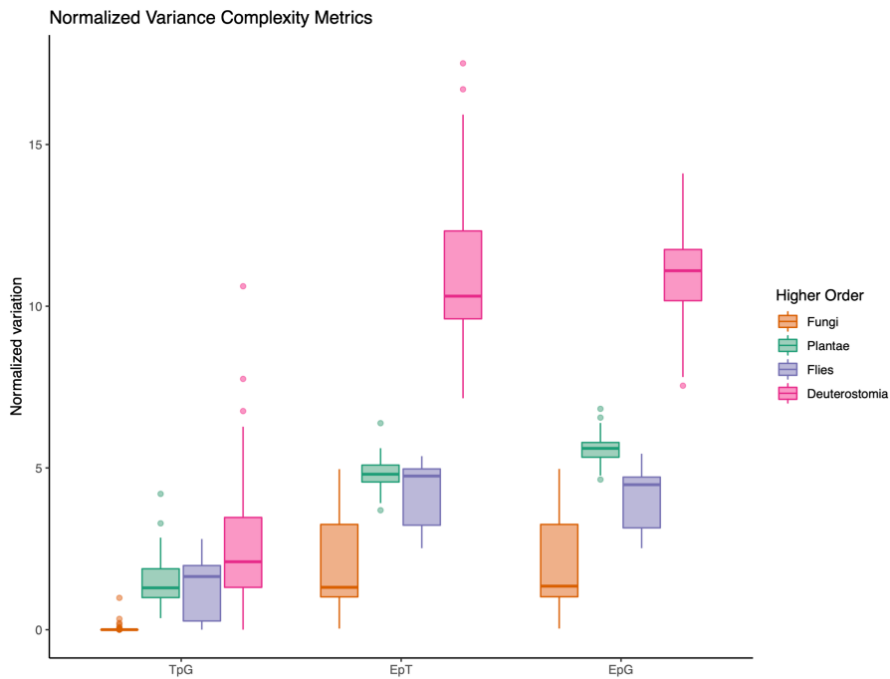

Supp. Figure 3: Normalized Variance Mean Complexity Metrics - Box and whisker plot of normalized variance computed from standard variance squared / mean from each complexity metric - TpG, EpT, and EpG. Higher-order taxonomic groups are colored by lineage. Outlier data are plotted as points above and below whiskers. Whiskers denote the extreme data, the box represents the upper and lower quartile, and box lines represent the median.

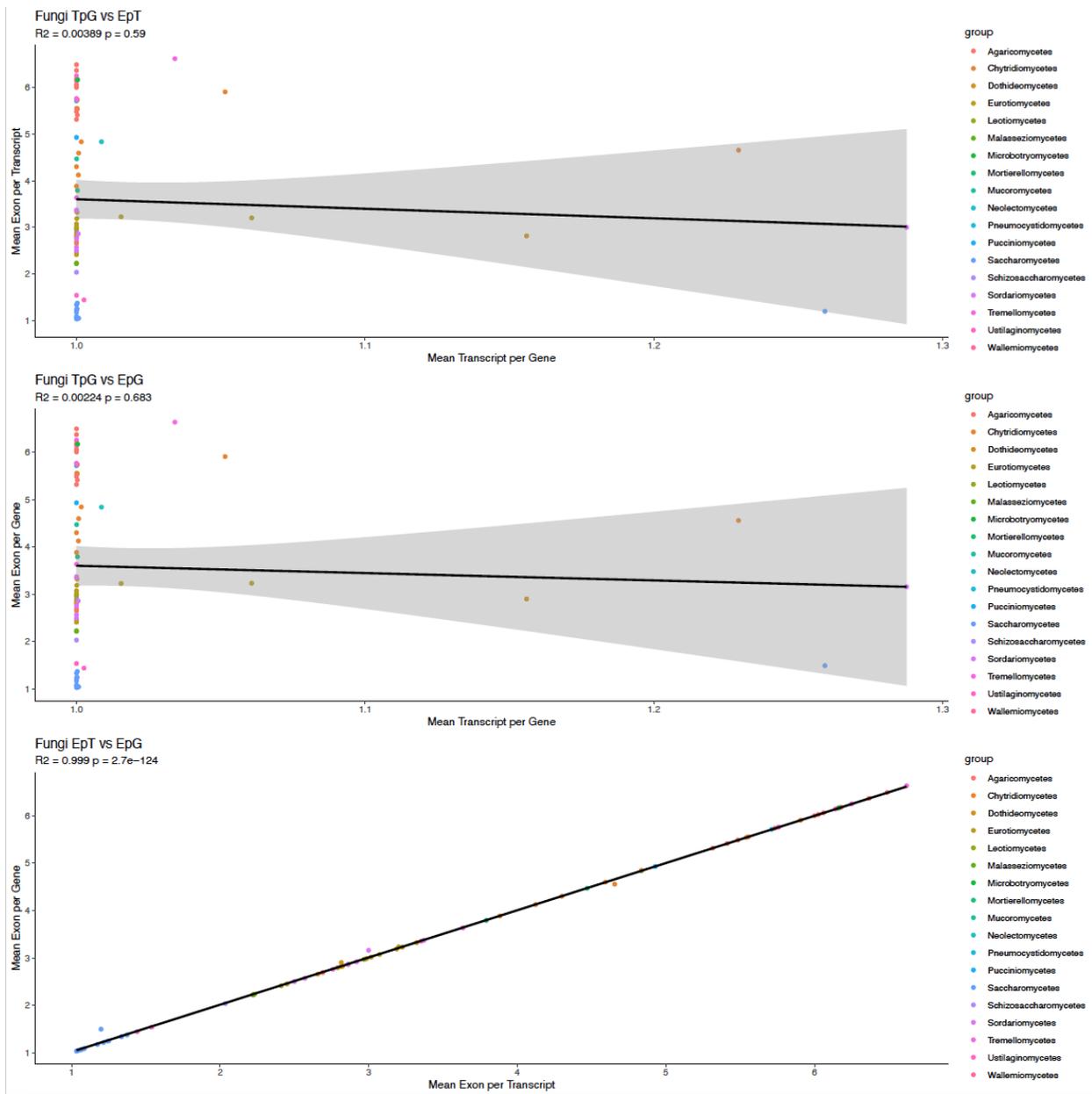

Supp. Figure 4: Fungi pair plots comparing complexity metrics. Metrics are compared among TpG, EpT, and EpG. The p-value and $R^2$ derived from calculating the Pearson correlation coefficient.

**COMPLEXITY CORRELATIONS**

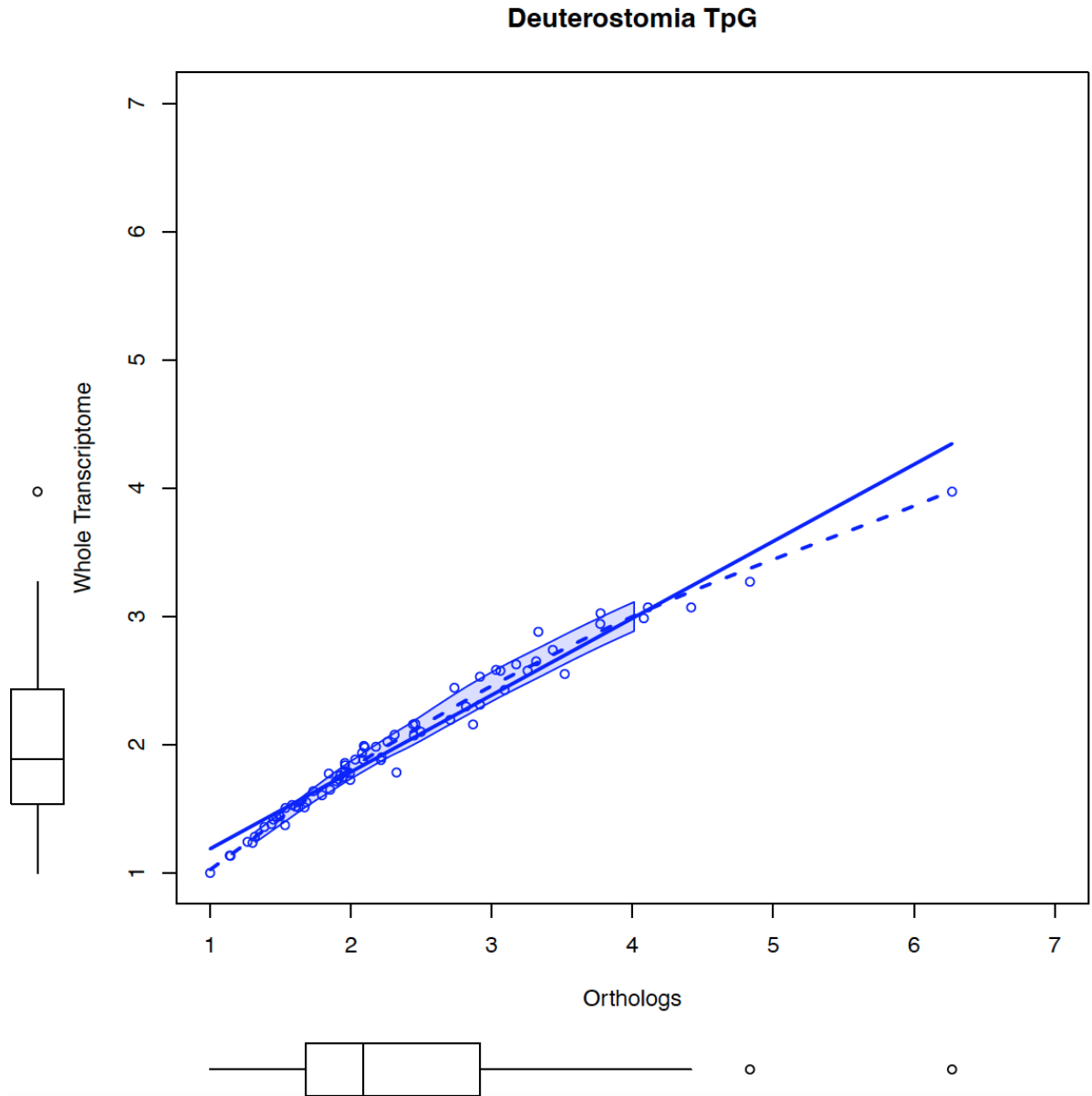

Supp. Figure 5: Deuterostome TpG scatter plot between whole-transcriptome (y-axis) and orthologs (x-axis) for each individual, the regression line (solid blue), the smoothed conditional spread (blue shaded regions) the non-parametric regression smooth (dotted blue line), and box and whisker plots for each dataset at the corresponding axes.

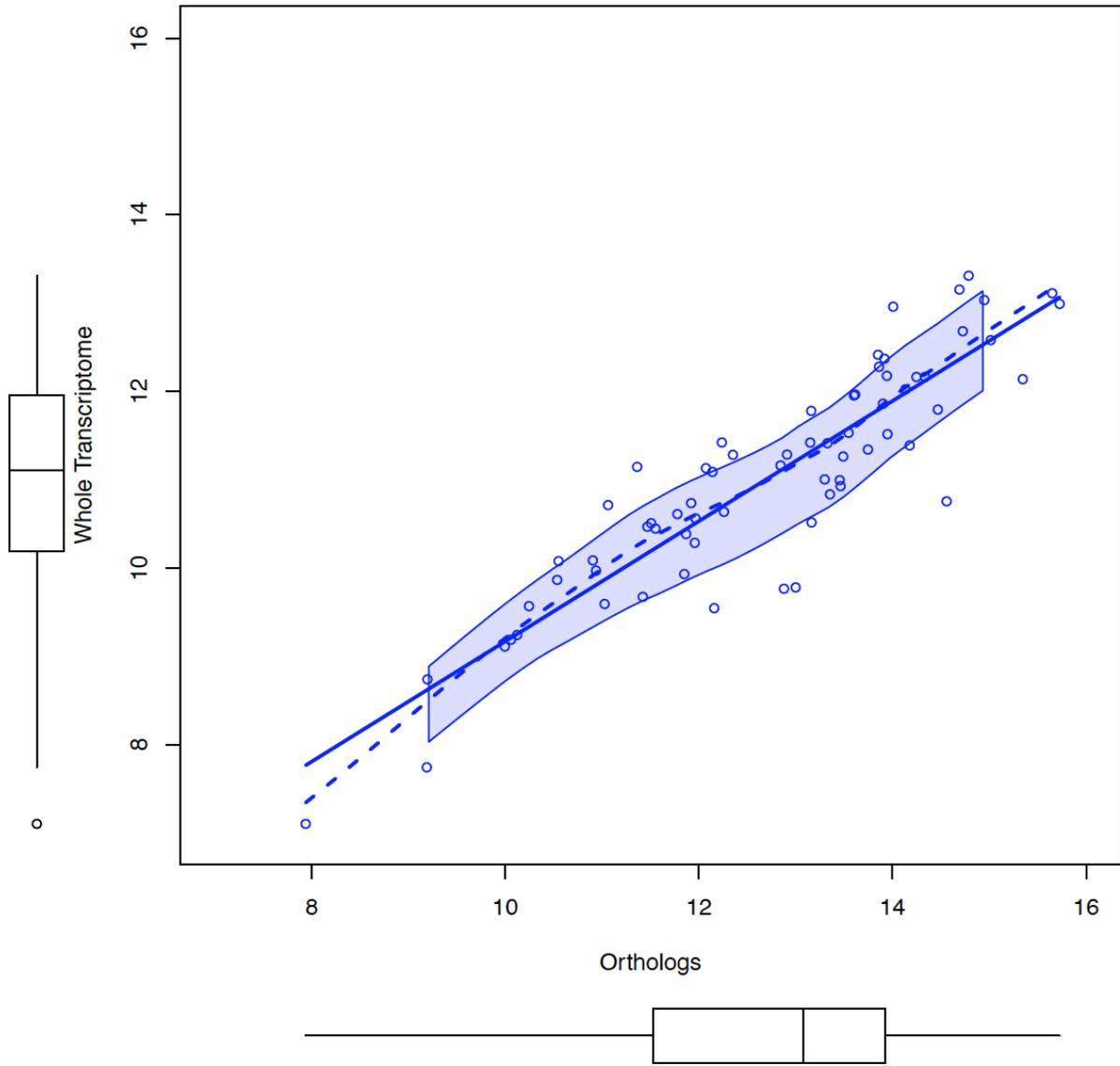

Supp. Figure 6: Deuterostome EpT scatter plot between whole-transcriptome (y-axis) and orthologs (x-axis) for each individual, the regression line (solid blue), the smoothed conditional spread (blue shaded regions) the non-parametric regression smooth (dotted blue line), and box and whisker plots for each dataset at the corresponding axes.

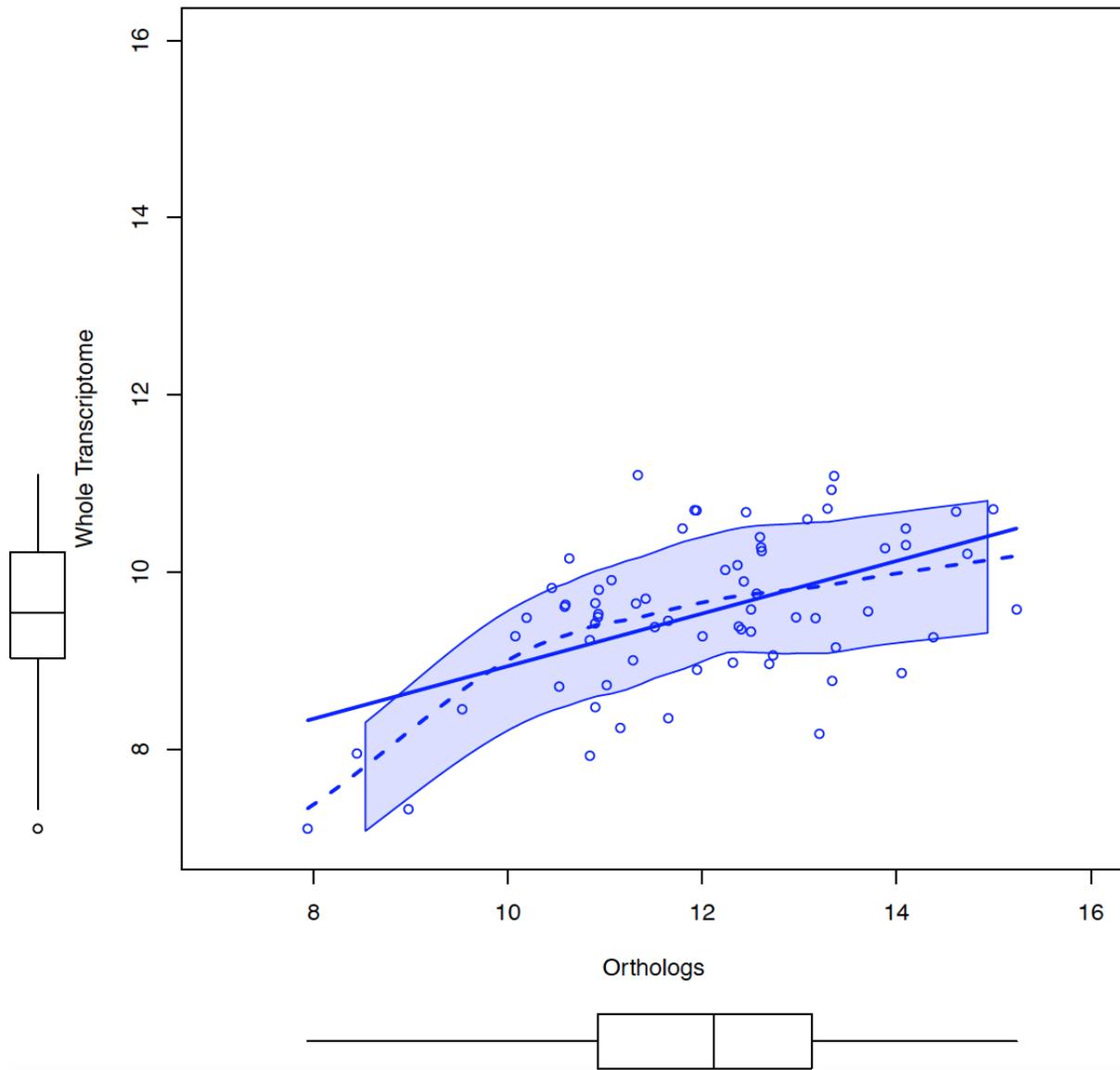

Supp. Figure 7: Deuterostome EpG scatter plot between whole-transcriptome (y-axis) and orthologs (x-axis) for each individual, the regression line (solid blue), the smoothed conditional spread (blue shaded regions) the non-parametric regression smooth (dotted blue line), and box and whisker plots for each dataset at the corresponding axes.

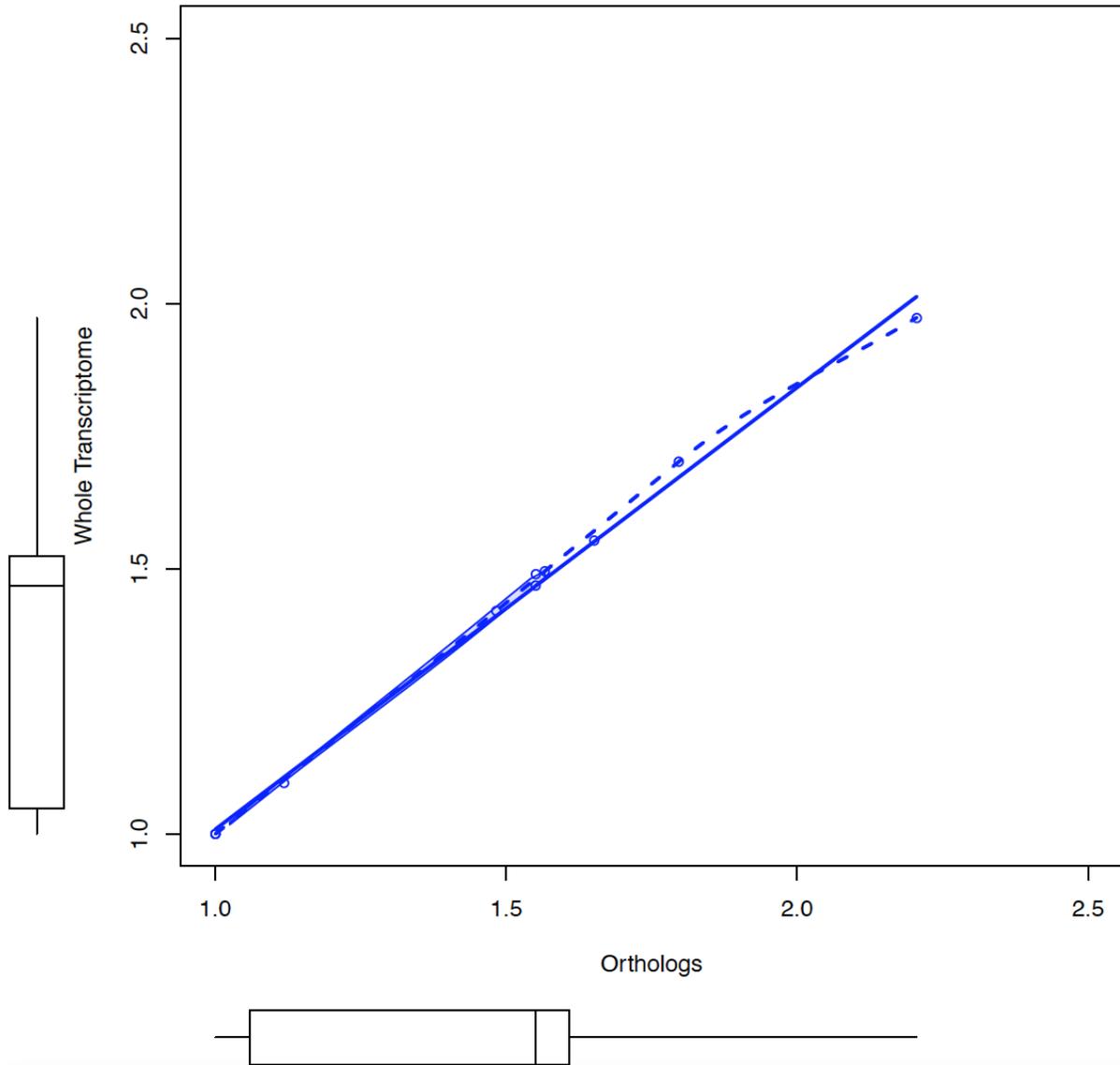

Supp. Figure 8: *Drosophila* TpG scatter plot between whole-transcriptome (y-axis) and orthologs (x-axis) for each individual, the regression line (solid blue), the smoothed conditional spread (blue shaded regions) the non-parametric regression smooth (dotted blue line), and box and whisker plots for each dataset at the corresponding axes.

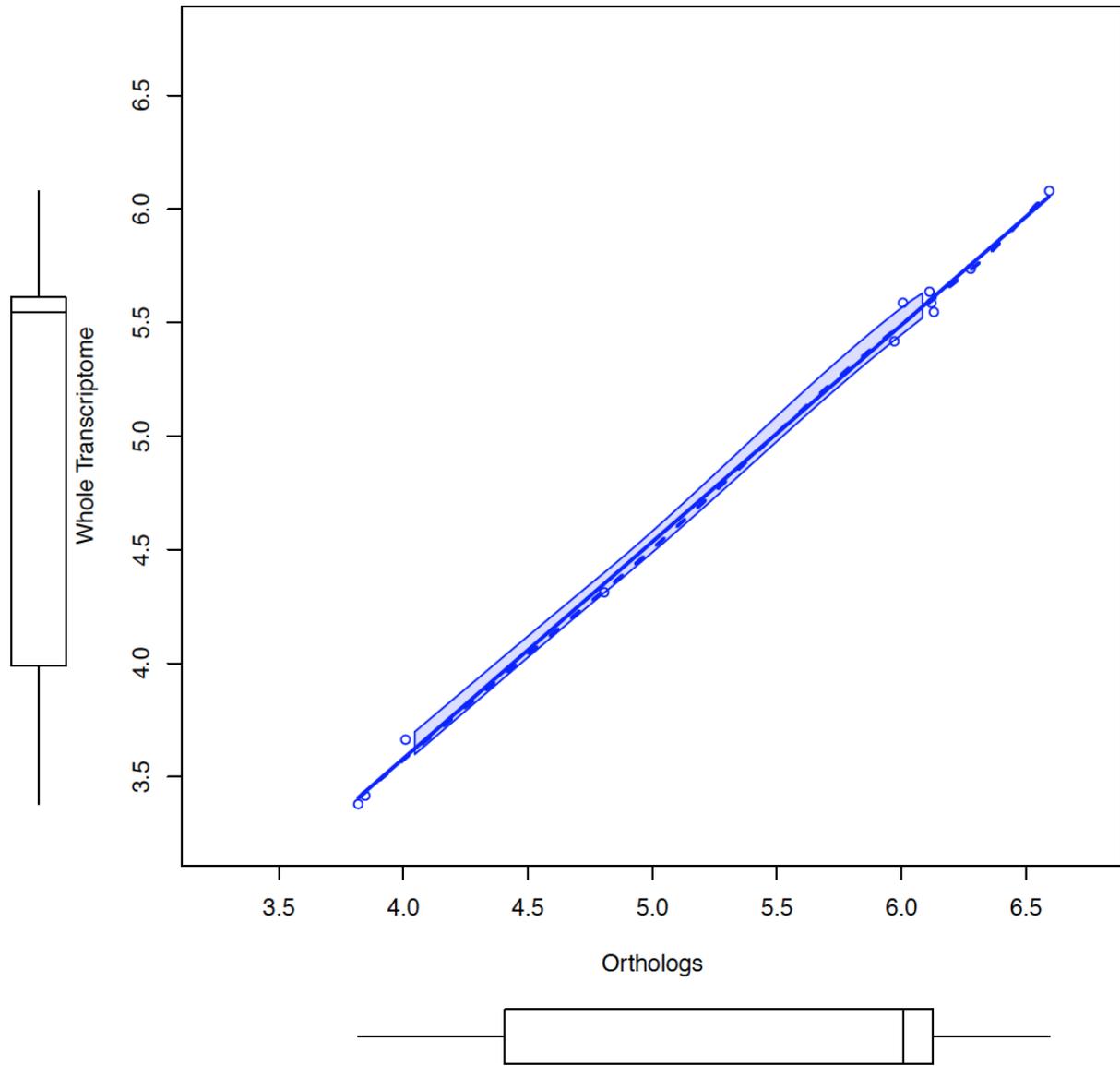

Supp. Figure 9: *Drosophila* EpT scatter plot between whole-transcriptome (y-axis) and orthologs (x-axis) for each individual, the regression line (solid blue), the smoothed conditional spread (blue shaded regions) the non-parametric regression smooth (dotted blue line), and box and whisker plots for each dataset at the corresponding axes.

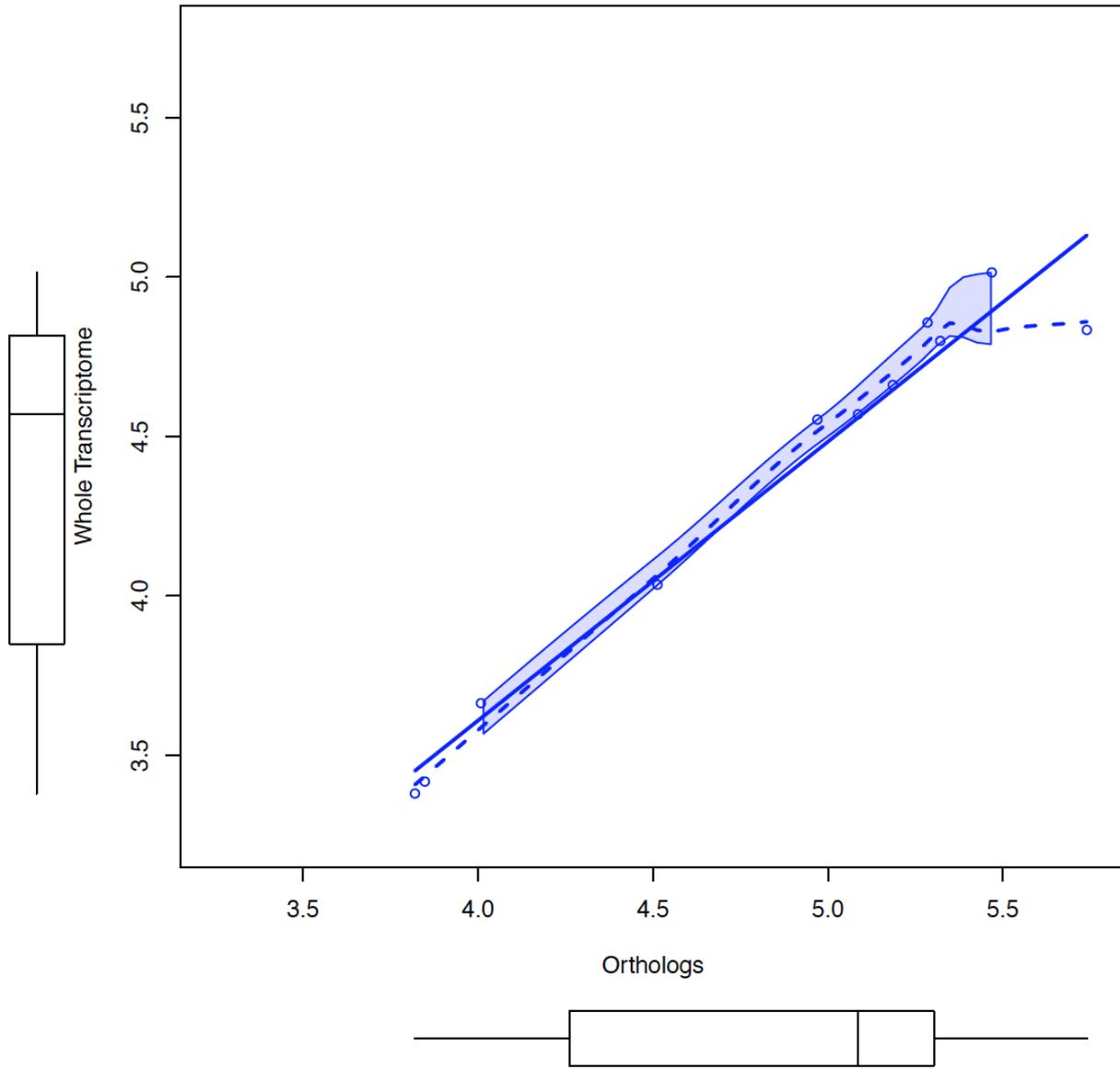

Supp. Figure 10: *Drosophila* EpG scatter plot between whole-transcriptome (y-axis) and orthologs (x-axis) for each individual, the regression line (solid blue), the smoothed conditional spread (blue shaded regions) the non-parametric regression smooth (dotted blue line), and box and whisker plots for each dataset at the corresponding axes.

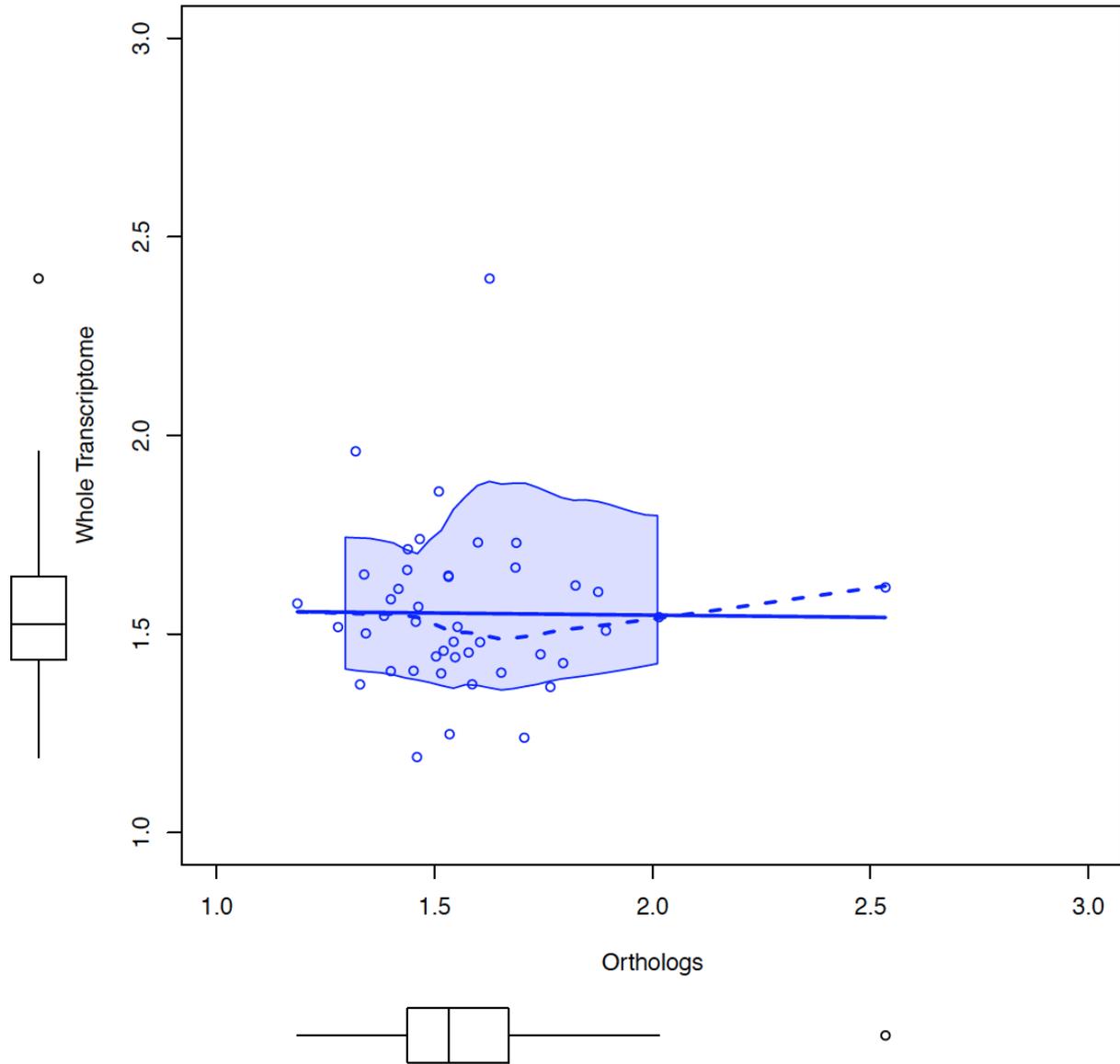

Supp. Figure 11: Plantae TpG scatter plot between whole-transcriptome (y-axis) and orthologs (x-axis) for each individual, the regression line (solid blue), the smoothed conditional spread (blue shaded regions) the non-parametric regression smooth (dotted blue line), and box and whisker plots for each dataset at the corresponding axes.

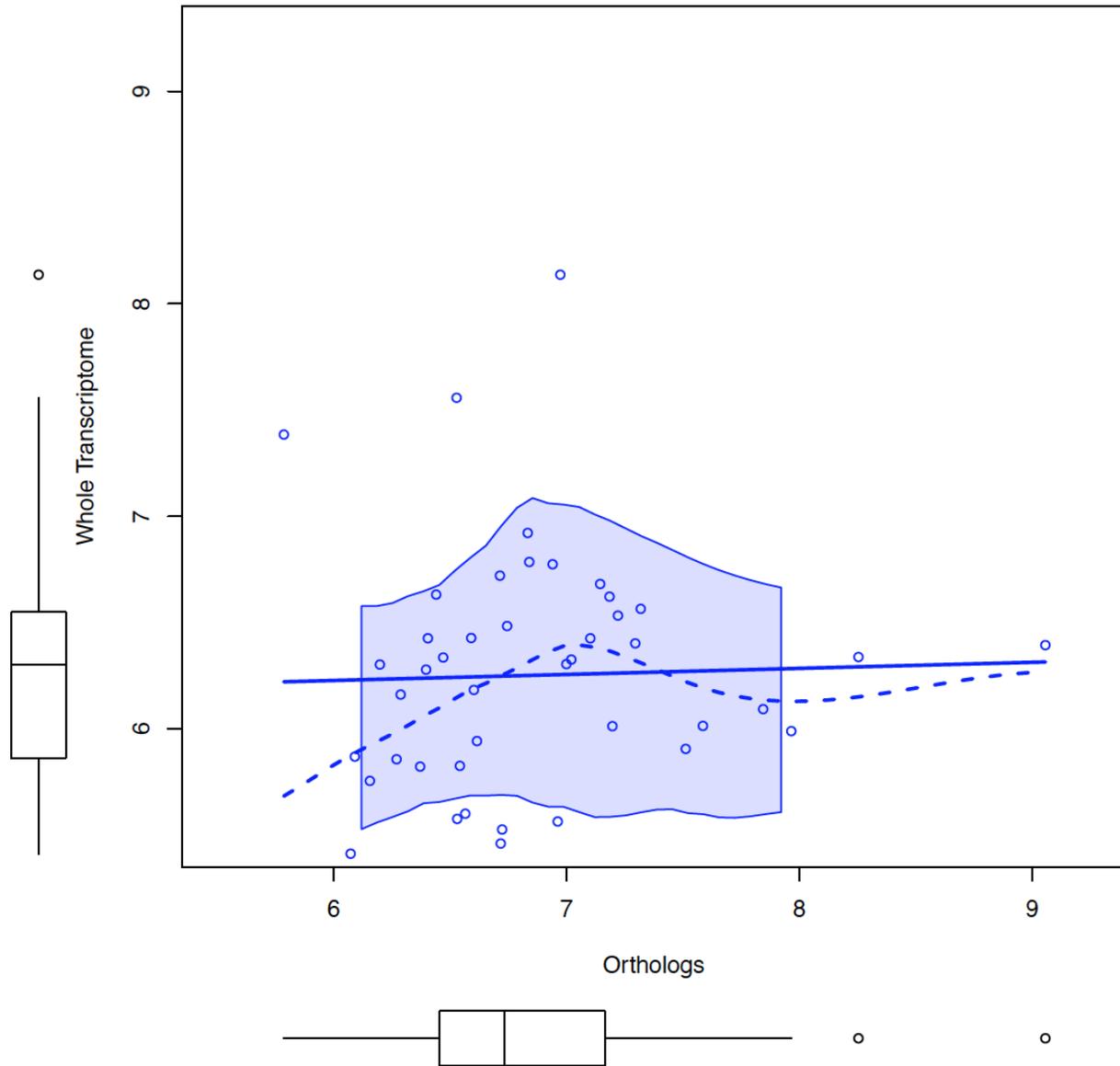

Supp. Figure 12: Plantae EpT scatter plot between whole-transcriptome (y-axis) and orthologs (x-axis) for each individual, the regression line (solid blue), the smoothed conditional spread (blue shaded regions) the non-parametric regression smooth (dotted blue line), and box and whisker plots for each dataset at the corresponding axes.

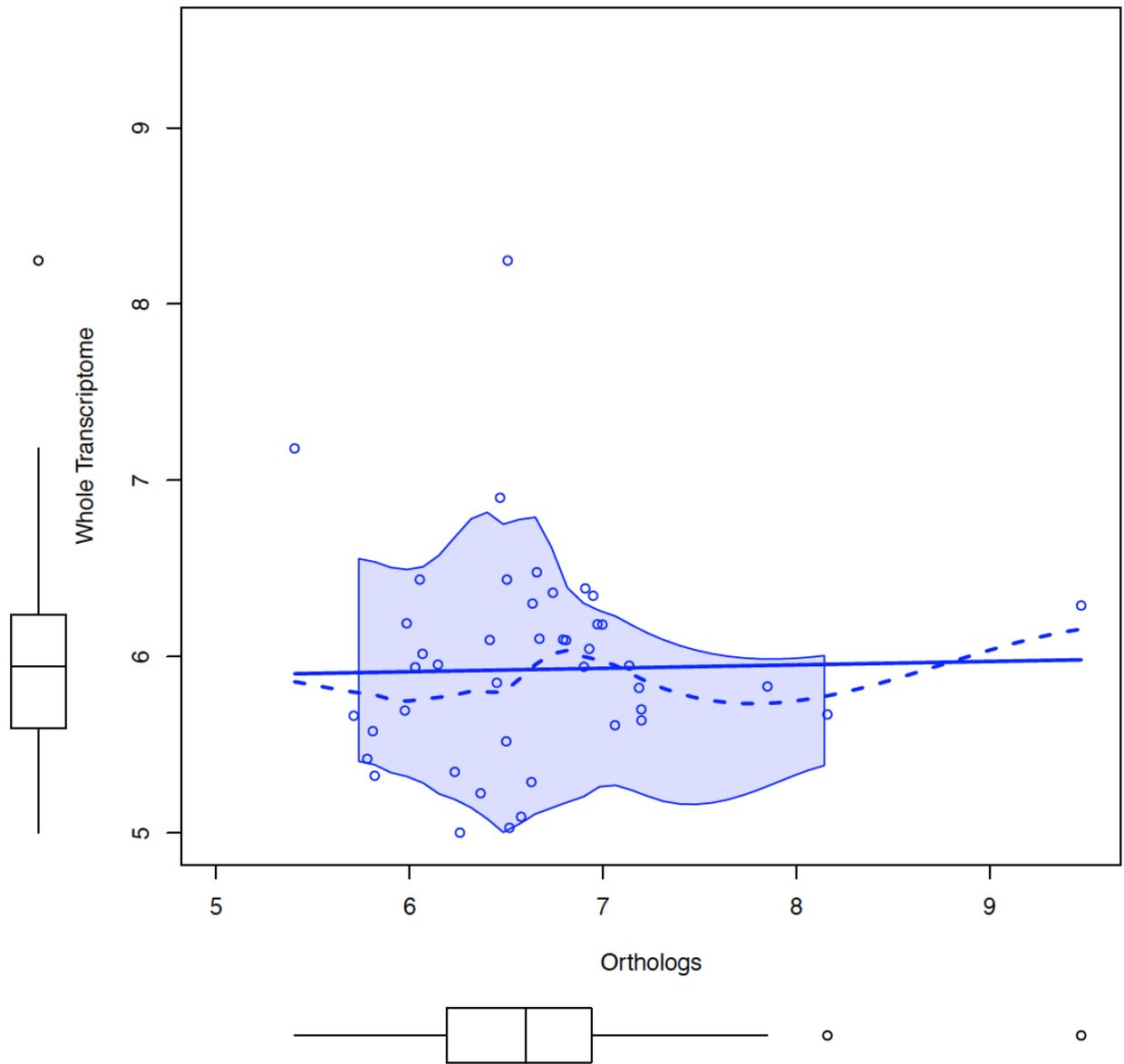

Supp. Figure 13: Plantae EpG scatter plot between whole-transcriptome (y-axis) and orthologs (x-axis) for each individual, the regression line (solid blue), the smoothed conditional spread (blue shaded regions) the non-parametric regression smooth (dotted blue line), and box and whisker plots for each dataset at the corresponding axes.

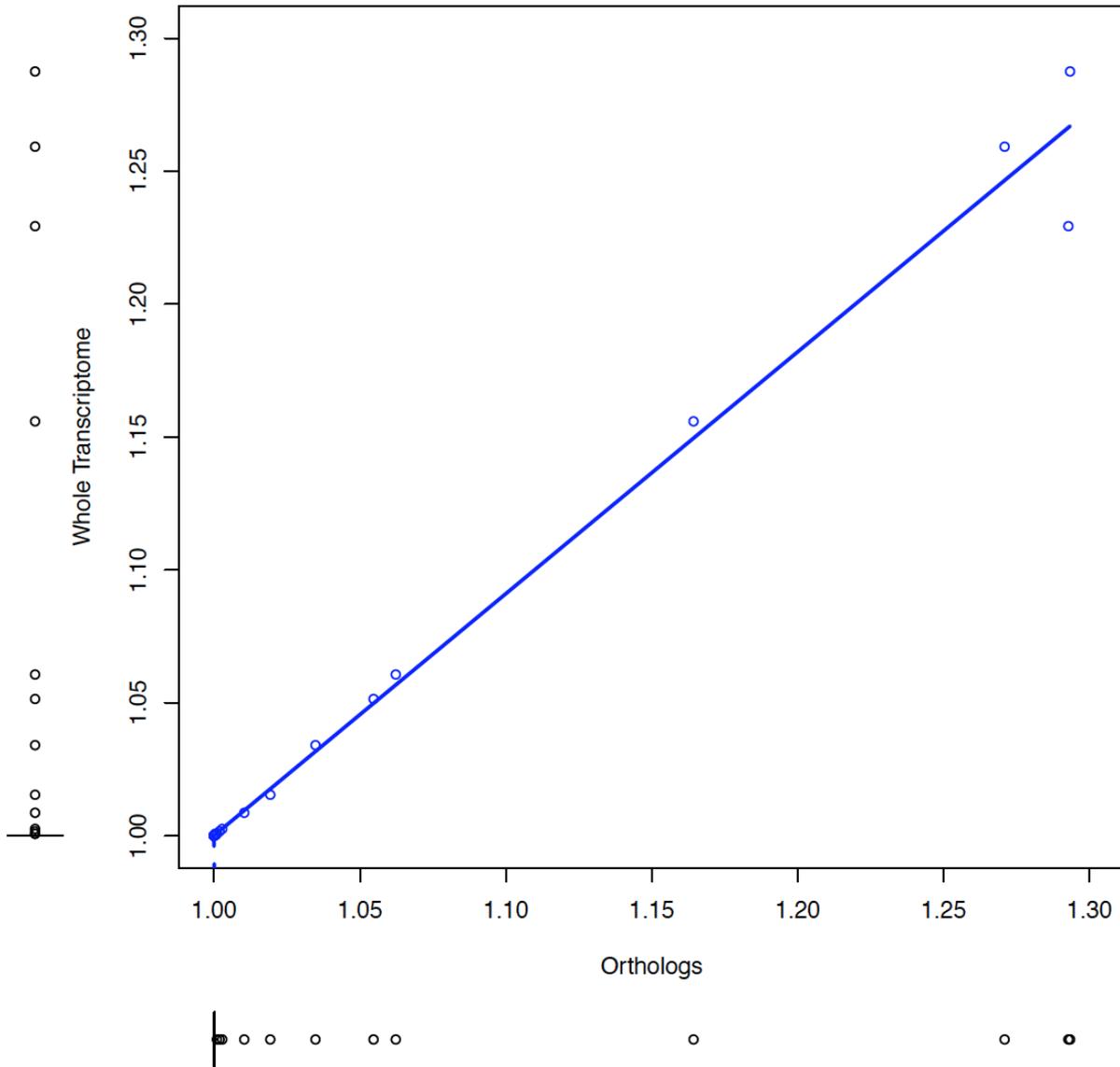

Supp. Figure 14: Fungi TpG scatter plot between whole-transcriptome (y-axis) and orthologs (x-axis) for each individual, the regression line (solid blue), the smoothed conditional spread (blue shaded regions) the non-parametric regression smooth (dotted blue line), and box and whisker plots for each dataset at the corresponding axes.

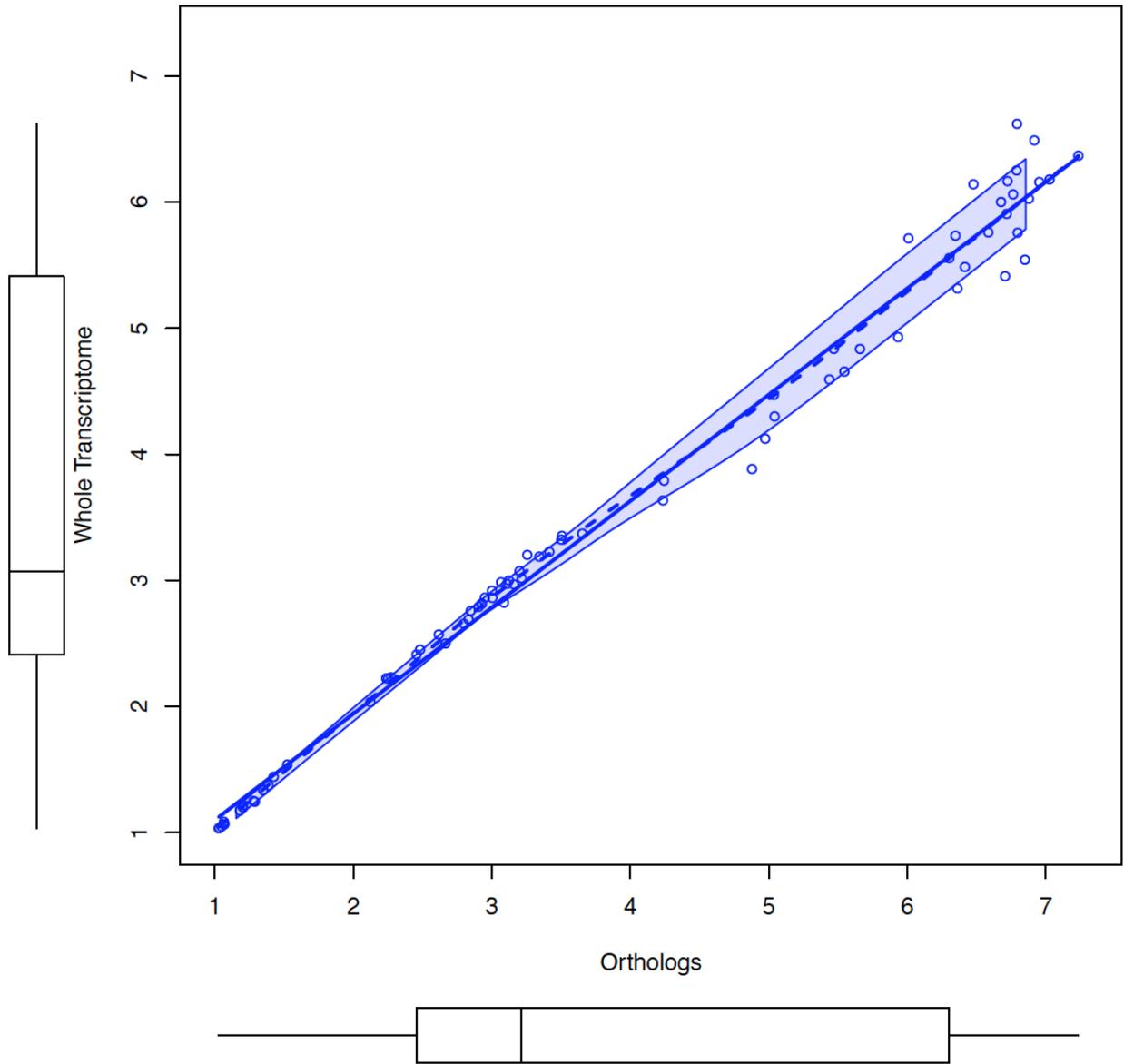

Supp. Figure 15: Fungi EpT scatter plot between whole-transcriptome (y-axis) and orthologs (x-axis) for each individual, the regression line (solid blue), the smoothed conditional spread (blue shaded regions) the non-parametric regression smooth (dotted blue line), and box and whisker plots for each dataset at the corresponding axes.

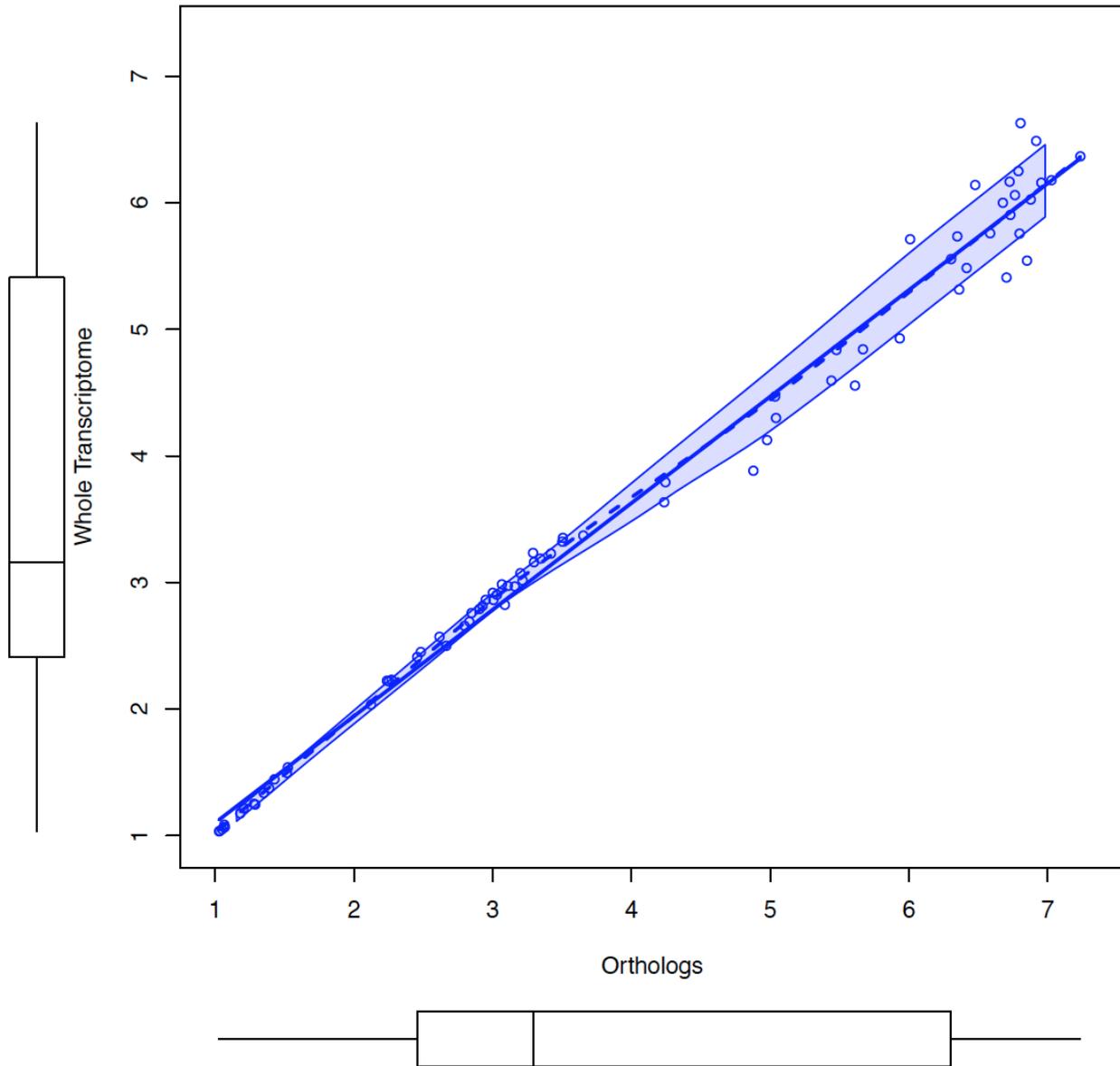

Supp. Figure 16: Fungi EpG scatter plot between whole-transcriptome (y-axis) and orthologs (x-axis) for each individual, the regression line (solid blue), the smoothed conditional spread (blue shaded regions) the non-parametric regression smooth (dotted blue line), and box and whisker plots for each dataset at the corresponding axes.

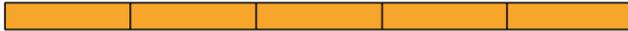

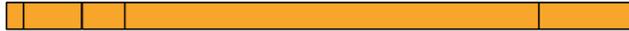

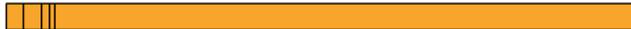

Supp. Figure 17: Effective Exon Number (EEN) captures information about the distribution of exon lengths produced when introns cause breaks in transcripts. When EEN = EpT, exon sizes are evenly distributed, more even than a model of random stochastic intron placement. If intron boundaries are drawn random across the transcript from a uniform distribution, they should follow the EEN distribution from a Broken Stick Model. More tightly clustered intron breakpoints yield lower EEN values compared with the Broken Stick. Patterns of variation in EEN are influenced by parameters of splicing models used during annotation, physical and molecular constraints on intron splicing processes, and constraints from natural selection shaping genetic structures.

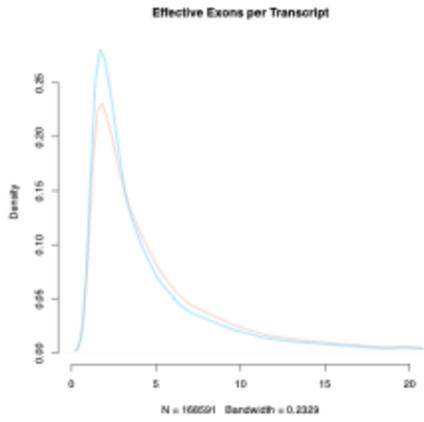

Supp. Figure 18: Effective exon number distribution for all genes in whole transcriptome data and conditioning on orthologs in humans. We observe a significant difference ($P<10^{-16}$) when lineage specific genes are excluded.

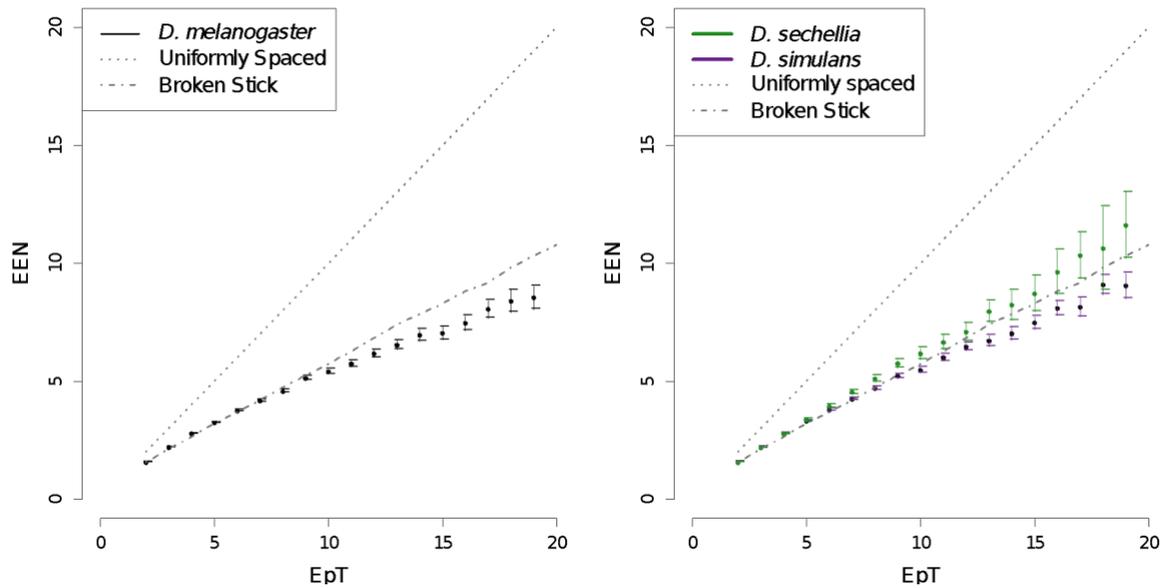

Supp. Figure 19: Mean EEN +/- 2*SE for A) *D. melanogaster* and B) two closely related species *D. simulans* and *D. sechellia.* In *D. melanogaster* EEN is consistent with the stick-breaking problem below 10 EpT. Above 10 EpT EEN is lower than expected, suggesting more tightly clustered introns than chance. *D. simulans* follows a similar pattern to *D. melanogaster.* In contrast, *D. sechellia* shows mean EEN greater than expected, but with wider error bars overlapping with the broken stick model at for higher EpT values. These different patterns for two very closely related species suggest differences in annotation, splicing mechanisms, or selective constraint influencing transcriptome complexity patterns. For some evolutionary comparisons, normalization for differences in complexity may be warranted.

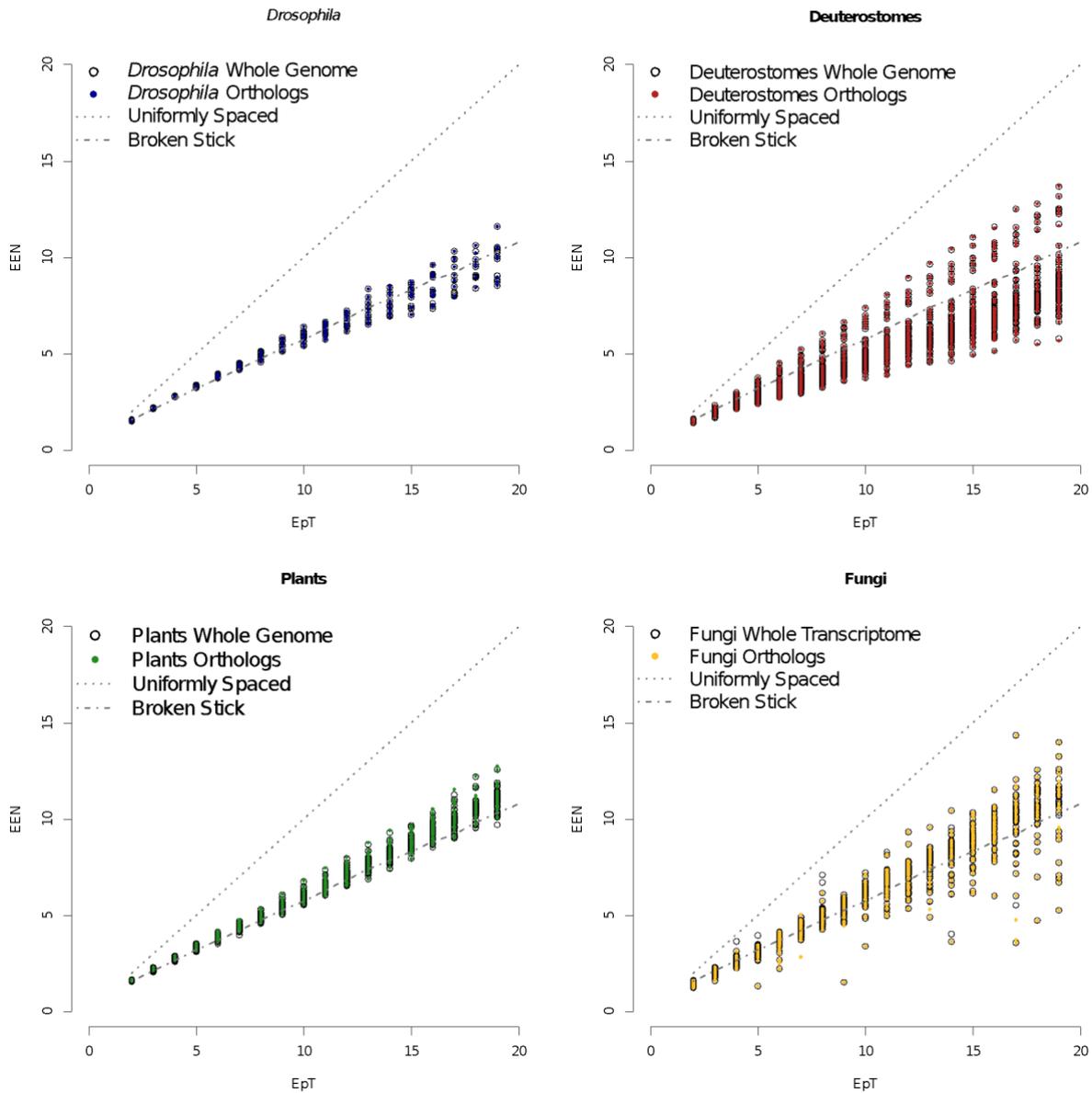

Supp. Figure 20: EEN vs EpT in Whole Transcriptome data and conditioning on Orthologs in A) Chordates and Deuterostomes and B) *Drosophila*. Conditioning on orthologs has only a nominal influence on EEN comparisons to the Broken Stick Model and does not alter inference for any species. The well validated annotations backed by abundant molecular evidence in humans (red) and *D. melanogaster* (yellow) show departures from the broken stick model. Both have reduced EEN for transcripts with high EpT suggesting more tightly clustered intron boundaries than expected under stochastic random processes. In chordates and deuterostomes, 62/68 species lie at or below the Broken Stick and 6 species lie above.

**EVOLUTIONARY RATES**

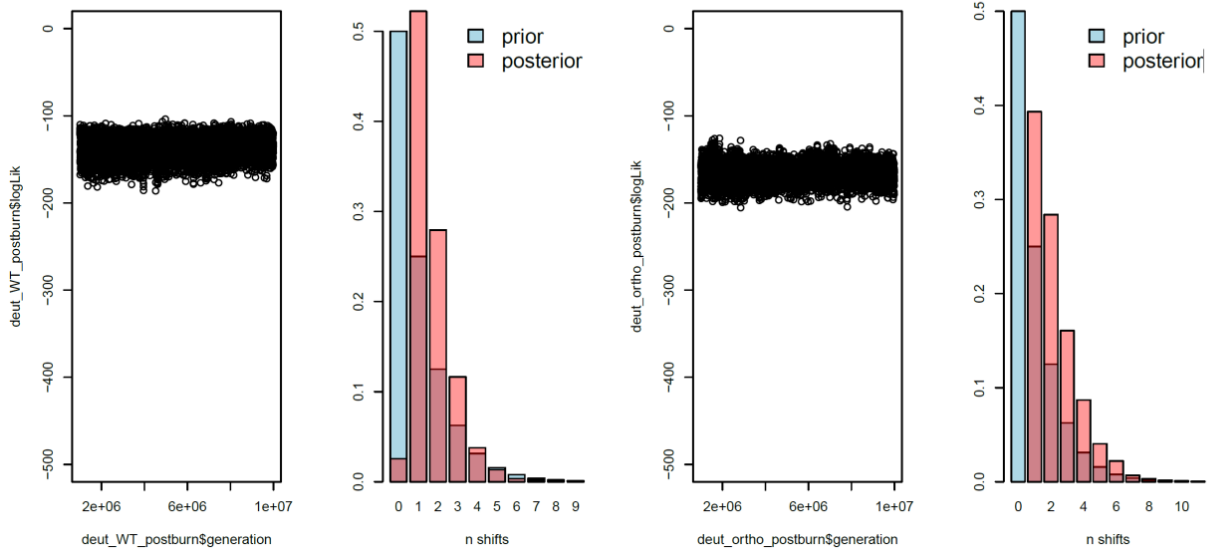

Supp. Figure 21: Deuterostome MCMC convergence post-burnin for whole-transcriptome (left) and ortholog datasets (right) with generation time on the x-axis and log-likelihood on the y-axis. Posterior probabilities for trait shifts for whole-transcriptome (left) and ortholog datasets (right), priors are in blue and posteriors are in red.

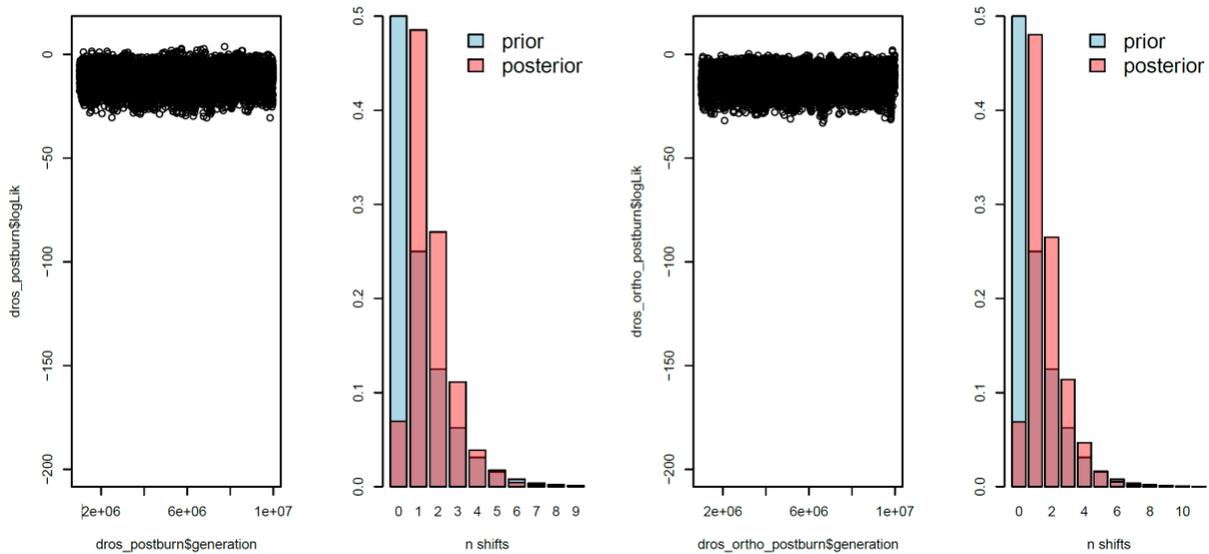

Supp. Figure 22: *Drosophila* MCMC convergence post-burnin for whole-transcriptome (left) and ortholog datasets (right) with generation time on the x-axis and log-likelihood on the y-axis. Posterior probabilities for trait shifts for whole-transcriptome (left) and ortholog datasets (right), priors are in blue and posteriors are in red.

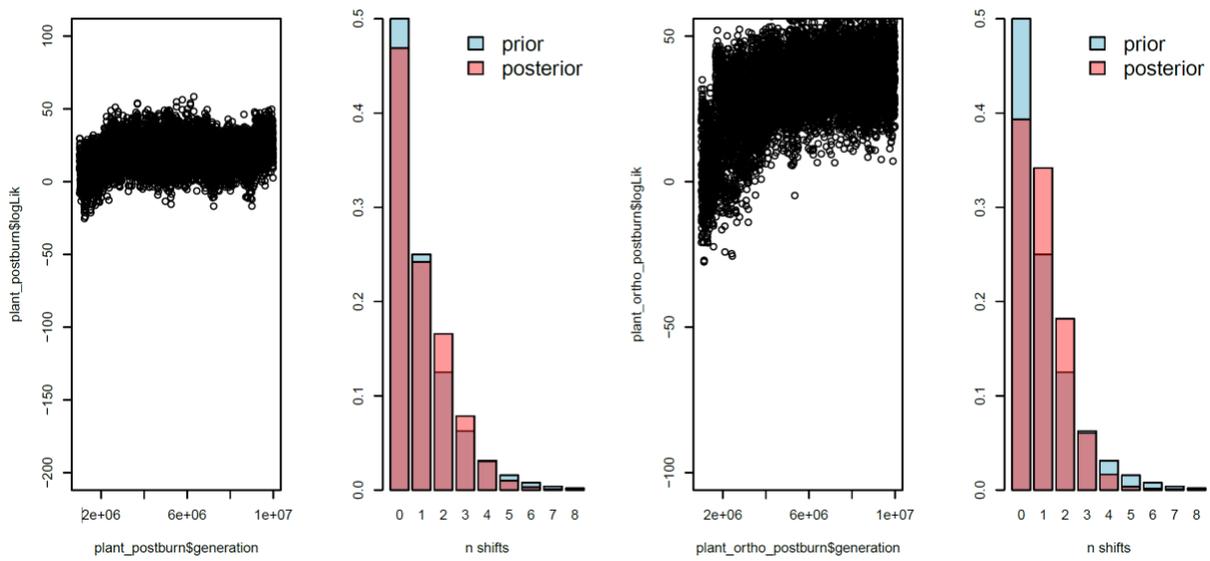

Supp. Figure 23: Plantae MCMC convergence post-burnin for whole-transcriptome (left) and ortholog datasets (right) with generation time on the x-axis and log-likelihood on the y-axis. Posterior probabilities for trait shifts for whole-transcriptome (left) and ortholog datasets (right), priors are in blue and posteriors are in red.

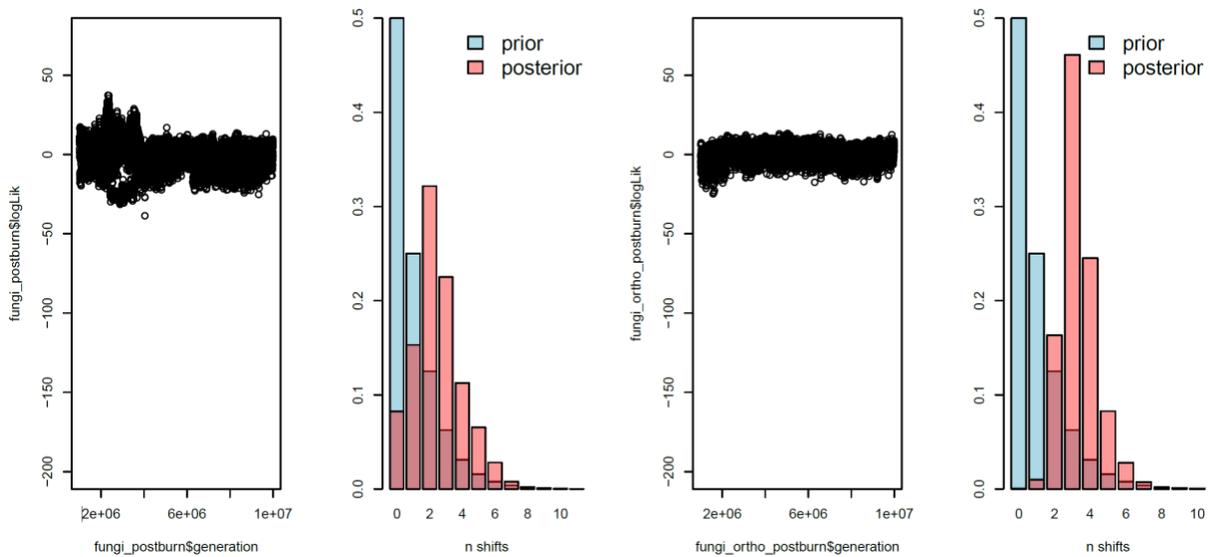

Supp. Figure 24: Fungi MCMC convergence post-burnin for whole-transcriptome (left) and ortholog datasets (right) with generation time on the x-axis and log-likelihood on the y-axis. Posterior probabilities for trait shifts for whole-transcriptome (left) and ortholog datasets (right), priors are in blue and posteriors are in red.